\documentclass[aps,prd,twocolumn,10pt,superscriptaddress,amsmath,amssymb,longbibliography]{revtex4-1}

\usepackage{graphicx}
\usepackage{subfigure}
\usepackage{epstopdf}
\usepackage{hyperref}
\usepackage{color}
\usepackage{bbold}

\newcommand{\magn}{\mathbf{m}}
\newcommand{\Magn}{\mathbf{M}}
\newcommand{\sigmab}{\boldsymbol{\sigma}}

\newcommand{\pos}{\mathbf{r}}

\newcommand{\mom}{\mathbf{k}}

\newcommand{\unitvec}{\hat{\mathbf{e}}}
\newcommand{\inpar}[1]{\left(#1\right)}
\newcommand{\inbra}[1]{\left[#1\right]}

\newcommand{\inang}[1]{\left<#1\right>}

\newcommand{\unity}{1\hspace{-0.27em}\text{\rmfamily l}}
\newcommand{\chern}{\mathcal{C}\!h}

\newcommand{\quotmarks}[1]{\textquotedblleft{#1}\textquotedblright}

\begin{document}

\title{Majorana modes in emergent-wire phases of helical and cycloidal magnet-superconductor hybrids}

\author{Stefan Rex}
\affiliation{Institute for Quantum Materials and Technologies, Karlsruhe Institute of Technology, 76021 Karlsruhe, Germany}
\affiliation{\mbox{Institut f\"ur Theorie der Kondensierten Materie, Karlsruhe Institute of Technology, 76128 Karlsruhe, Germany}}
\author{Igor V. Gornyi}
\affiliation{Institute for Quantum Materials and Technologies, Karlsruhe Institute of Technology, 76021 Karlsruhe, Germany}
\affiliation{\mbox{Institut f\"ur Theorie der Kondensierten Materie, Karlsruhe Institute of Technology, 76128 Karlsruhe, Germany}}
\affiliation{Ioffe Institute, 194021 St. Petersburg, Russia}
\author{Alexander D. Mirlin}
\affiliation{Institute for Quantum Materials and Technologies, Karlsruhe Institute of Technology, 76021 Karlsruhe, Germany}
\affiliation{\mbox{Institut f\"ur Theorie der Kondensierten Materie, Karlsruhe Institute of Technology, 76128 Karlsruhe, Germany}}
\affiliation{L. D. Landau Institute for Theoretical Physics RAS, 119334 Moscow, Russia}
\affiliation{Petersburg Nuclear Physics Institute, 188300 St. Petersburg, Russia}

\begin{abstract}
Noncollinear magnetism opens exciting possibilities to generate topological superconductivity. Here, we focus on helical and cycloidal magnetic textures in magnet-superconductor hybrid structures in a background magnetic field. We demonstrate that this system can enter a topological phase which can be understood as a set of parallel topological wires. We explore and confirm this idea in depth with three different approaches: a continuum model, a tight-binding model based on the magnetic unit cell, and exact diagonalization on a finite two-dimensional lattice. The key signature of this topological state is the presence of Majorana bound states at certain disclination defects in the magnetic texture. Based on the $C_2$ symmetry imposed by the helical or cycloidal texture, we employ the theory of topological crystalline superconductors with rotation invariants to obtain the Majorana parity at disclinations. Furthermore, we consider a 90-degree helimagnet domain wall, which is formed by a string of alternating disclinations. We discuss how the resulting chain of disclination bound states hybridizes into two chiral modes with different velocities. We suggest that hybrid systems of chiral magnets and superconductors are capable of hosting Majorana modes in various spatial configurations with potentially far less nano-engineering than in, e.g., semiconductor wires.
\end{abstract}

\maketitle

\section{Introduction}\label{Sec:Intro}
Since the first predictions of self-conjugate quasiparticles in topological superconductors \cite{Kit01, Iva01}, tremendous research activity aimed at turning such Majorana modes into a physical reality has been taking place. The interest in Majorana physics is partially fundamental but also fueled by envisioned applications in topological quantum computing. Namely, non-commutative exchange statistics is expected for Majorana modes in two dimensions \cite{NSS08,Ali12,Ben13,StL13}.

For several years, efforts toward Majorana bound states (MBS) in superconducting systems were mostly centered around semiconductor nanowires \cite{Kit01,LSD10,ORO10} with superconducting substrates or coating. Experimentally, the first signatures of MBS were measured in such systems \cite{MZF12, DYH12, DRM12, RLF12}, and significant progress has been reported henceforth \cite{DVH16,GZB18,VWH20}. However, intricate nanowire setups do not allow for much flexibility, e.g., with respect to braiding.

Subsequently, various magnet-superconductor hybrids (MSH) have been discussed as a promising alternative framework for investigating Majorana physics, including arrays \cite{CEA11,NDB13,PGO13,KSY13,BrS13,VaF13,NTN13,HKS14,PWR14,HMK15,BSH15,RoO15,PPG15,CSF16,SFC16,LNW16,KPO20} or islands \cite{ChS15,MCR19,SAK20,PAG20} of magnetic adatoms on a superconducting substrate and inhomogeneous external or intrinsic fields \cite{KWF12,KSL12,MaM12,LuW13,KlL13,SAB15,FMS16,MSK17,ZMH19,VZS18,VAZ19,KSP20}. Many of the proposed systems allow for topological states in two dimensions. In ferromagnetic islands, signatures of chiral Majorana edge modes have already been observed in experiments \cite{MGB17,MMB19,PMC19}. Furthermore, intriguing effects are expected for spatially varying exchange fields, because they contribute to the effective spin-orbit coupling (SOC) \cite{BJK10, CEA11, KWF12}. Self-organized noncollinear magnetic textures are found in chiral magnets, which are therefore promising constituents for MSH. More precisely, the Dzyaloshinskii-Moriya interaction \cite{Dzy58,Mor60} at interfaces can lead to spin spirals, skyrmion crystals, free skyrmions, and several further textures \cite{HFT17}. Promising candidate systems showing chiral magnetism on substrates capable of superconductivity have been identified in experiments \cite{HDL18, KBW20}.

The vector field of the local magnetic moments carries topological information. It is well-known that magnetic skyrmions are characterized by an integer topological charge \cite{NaT13}, but topological features are found in, e.g., helimagnets as well \cite{SMK18}. Interestingly, in MSH there appear to be synergy effects between topological defects of the magnetic texture on the one hand, and topologically protected bound states of the superconductor on the other. For the case of skyrmions, any even topological charge has been shown to correspond to the emergence of a MBS \cite{YSK16,RGM19,GMS19} at the skyrmion center, whereas the Majorana parity is flipped when a vortex carrying an odd number of flux quanta is added to the system \cite{RGM19}, as in a skyrmion-vortex pair \cite{HSR16,DVE18,BCB19}.

\begin{figure}[t]
\includegraphics[width=0.93\columnwidth]{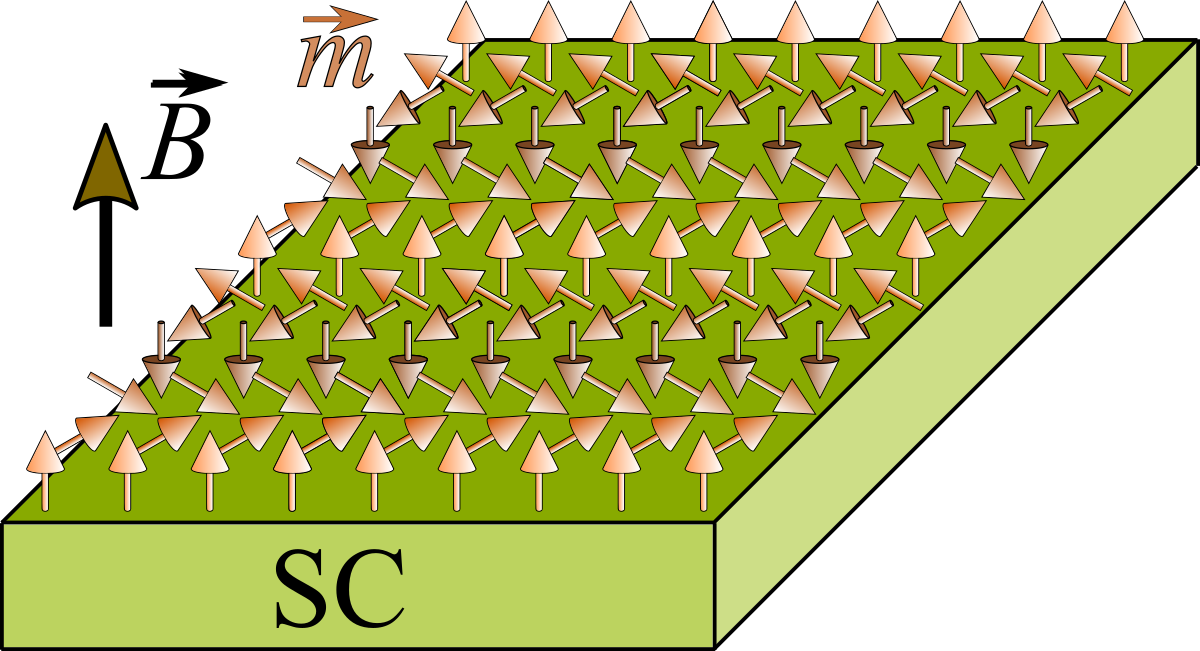}
\caption{\label{Fig:Sketch}A sketch of the hybrid system consisting of a superconducting layer covered by a chiral magnet in a uniform background field. The magnetization rotates along one axis, where the spin spiral may be helical or cycloidal.}
\end{figure}

In one theoretical proposal \cite{GSK18}, it was pointed out that elongated skyrmions can effectively turn into Majorana wires with MBS at both ends. Interestingly, an elongated skyrmion can be thought of as a (finite-length) 360-degree domain wall. A 360-degree wall, on the other hand, is topologically equivalent to a stripe of a helical or cycloidal magnet. Thus, it is plausible that helical or cycloidal MSH, as illustrated in Fig.~\ref{Fig:Sketch}, may have a phase in which they form a collection of parallel Majorana wires. Adding defects to the magnetic pattern may then lead to networks of wires, with additional MBS at ends or odd junctions of wires.

The aim of this paper is to work out this idea in more detail and establish a theoretical foundation for the phase of emergent Majorana wires at the interface of a superconductor and a chiral magnet with spiral order. We will start in Sec.~\ref{Sec:Continuum} from a simple continuum model of the superconductor under the assumption that the net exchange field, which stems from the magnetic layer and a uniform external field, varies slowly in space. This point of view allows for valuable analytical insight into the conditions that need to be met for the topological phase to appear, although it will not be quantitatively exact except for very large magnetic period lengths. The basic prerequisites for topological wires are known from semiconductor heterostructures \cite{LSD10,ORO10}: confinement to (quasi-)one dimension, SOC, Zeeman splitting, and an $s$-wave pairing gap. We will shed light on how these conditions translate into our MSH. Some of the results from the section dealing with the continuum model have already been reported in previous studies. Nevertheless, we attempt to give a complete survey of the properties of this model.

In Sec.~\ref{Sec:C2TB}, we turn to a tight-binding model based on the magnetic unit cell. The magnetic texture is $C_2$ symmetric, and we employ this symmetry to create a link to topological crystalline superconductors with rotation invariants \cite{TeH13, GTH13, BTH14}. This theoretical framework readily provides us with a topological classification of defects, namely $\pm\pi$ disclinations, in the system. We study the phase diagram and identify a sizable nontrivial phase that is consistent with the continuum picture of emergent Majorana wires.

In Sec.~\ref{Sec:ExDiag}, we confirm our results on MBS at disclinations by means of exact diagonalization on a finite two-dimensional grid. As an example of a system with multiple disclinations, we discuss in Sec.~\ref{Sec:DomainWall} a 90-degree domain wall within the magnetic texture, which is known to consist of a chain of disclinations \cite{SMK18}. For this case, we derive how the MBS can hybridize into two counterpropagating chiral modes between the domains. Finally, we summarize our results in Sec.~\ref{Sec:Conclusion}.

\section{Continuum model}\label{Sec:Continuum}
Here we consider the MSH as a nearly free electron system in two dimensions. Most of this section will rely on the assumption of slowly varying fields, i.e., the spin spiral period length is large compared to the atomic lattice constant. In this limit, several useful properties can be calculated exactly.

We start from the Bogoliubov--de Gennes (BdG) Hamiltonian in real space, $\pos=(x,y)$, using the spinor basis $\inpar{c_{\pos\uparrow},c_{\pos\downarrow},c^\dagger_{\pos\uparrow},c^\dagger_{\pos\downarrow}}$, where we denote Pauli matrices by $\sigma$ when acting in spin space, and by $\tau$ when acting in Nambu (particle-hole) space. For this choice of basis, it is useful to define $\tilde{\sigmab}=(\sigma_x,\sigma_y\tau_z,\sigma_z)$. 
The BdG Hamiltonian then reads
\begin{eqnarray}
\hat{H}(\pos) &=&
\inbra{-\frac{\hbar^2\nabla^2}{2m}-\mu
+ \Magn(\pos)\cdot\tilde{\sigmab}}\tau_z \notag\\
&&{}- i\hbar\alpha\unitvec_z\cdot\inpar{\tilde{\sigmab}\times\nabla}
- \Delta\sigma_y\tau_y \,.  \label{Eq:BdGRealSpace}
\end{eqnarray}

It consists of the kinetic term (neglecting orbital effects) with effective electron mass $m$, the chemical potential $\mu$, the exchange field $\Magn(\pos)$, Rashba SOC of strength $\alpha$, and $s$-wave singlet superconducting pairing with an amplitude $\Delta$, which is chosen to be real.

The exchange field comprises the spatially varying magnetization $\magn(\pos)$ as well as a uniform background field $\mathbf{B}=B\unitvec_z$ perpendicular to the plane,
\begin{equation}\label{Eq:ExField}
\Magn(\pos) = B\unitvec_z+J\magn\inpar{\pos} \,.
\end{equation}
The Zeeman factor $\frac12g_e\mu_B$ has been absorbed into $B$, and $J$ is a coupling constant arising from the proximity of the magnetic layer to the superconductor. Thus, $\Magn$ has units of energy.

We align the coordinate axes such that the spiraling magnetic texture depends only on $y$,
\begin{equation}\label{Eq:MagTexture}
  \magn(\pos) = \magn(y) = |\magn|\inbra{\cos\inpar{\frac{2\pi y}{l}}\unitvec_z 
	                                     + \sin\inpar{\frac{2\pi y}{l}}\unitvec_j} ,
\end{equation}
where $j=x$ for helical magnets (Bloch-like rotation) and $j=y$ for out-of-plane cycloidal magnets (N\'eel-like rotation). We will use the spherical coordinates $M(y), f, g(y)$ for the net field $\Magn$,
\begin{equation}\label{Eq:MShape}
\Magn = M\inbra{\cos(f) \sin(g) \unitvec_x + \sin(f) \sin(g) \unitvec_y + \cos(g) \unitvec_z} ,
\end{equation}
where it follows from Eqs.~\eqref{Eq:ExField} and \eqref{Eq:MagTexture} that
\begin{equation}\label{Eq:gAngle}
\tan g(y) = \frac{\sin\inpar{2\pi y/l}}{\cos\inpar{2\pi y/l} + B/(J|\magn|)} \,.
\end{equation}
For the other angle, we simply have $f\equiv 0$ for helical magnets and $f\equiv \frac12\pi$ for cycloidal magnets.

\subsection{Spin-alignment transformation}\label{Sec:SAT}
As a first step, we perform a spin-alignment transformation (SAT) on the system, by which $\Magn(\pos)$ is locally rotated to the positive $z$ direction. The SAT has become a standard tool in the analysis of nonuniform magnets \cite{Vol87, BJZ98, BaM07, BJK10, ZMH11, KWF12}. In Nambu space, it is expressed through the position-dependent unitary matrix
\begin{equation}
\hat{U}(\pos) =
\begin{pmatrix}
  U(\pos)   &   0    \\
	0   &  \sigma_x U^\dagger(\pos)\sigma_x
\end{pmatrix}
\end{equation}
with the spin-space rotation matrix
\begin{equation}\label{Eq:SpinRotation}
U(\pos) = \cos\frac{g(\pos)}{2} + i\inbra{\sigma_x\sin f(\pos) - \sigma_y\cos f(\pos)}\sin\frac{g(\pos)}{2}.
\end{equation}

Subsequent to the SAT, we take the Hamiltonian to momentum space. However, assuming that the original field $\Magn(\pos)$ varies sufficiently slowly, we can keep the position $\pos$ as a parameter. %
The entire transformed local BdG Hamiltonian then reads
\begin{eqnarray}\label{Eq:HcontSAT}
\hat{\tilde H}(\mom;\pos) &=& \inbra{\frac{\hbar^2k^2}{2m} -\mu_\text{eff}(\pos) + M\!(\pos)\sigma_z}\tau_z \notag\\&&{}+ \hbar \mathbf{s}(\mom;\pos) \cdot \tilde{\sigmab} - \Delta\sigma_y\tau_y
\end{eqnarray}
in the momentum-space spinor basis $\inpar{c_{\mom\uparrow},c_{\mom\downarrow},c^\dagger_{-\mom\uparrow},c^\dagger_{-\mom\downarrow}}$. The chemical potential and the SOC term are modified by the SAT, and the resulting quantities $\mu_\text{eff}$ and $\mathbf{s}$ will be discussed in the following subsections. The pairing term, on the other hand, is not affected by the SAT, because singlet Cooper pairs do not carry spin.

Expressions for the transformation of the individual terms in $\hat{H}$ for general $f(\pos)$ and $g(\pos)$ are provided in Appendix~\ref{App:SAT}. In what follows, we only consider spiraling magnetic patterns according to Eq.~\eqref{Eq:MagTexture}.

The SAT is not unique. Namely, the process of rotating $\Magn$ onto the $z$ axis could be interrupted at any stage by additional spin-space rotations around the $z$ axis (with smoothly varying angles) without changing the final orientation of $\Magn$. In general, this equips the transformed Hamiltonian with an effective $SU(2)$ gauge freedom. It can be reduced to an $U(1)$ gauge theory in the adiabatic limit, where it has been dubbed \emph{emergent} electrody\-na\-mics \cite{Vol87, BJZ98, BaM07, ZMH11, SRB12}. The terms generated by the transformation are not individually gauge-independent. Only in their entirety can they be written in a covariant manner. However, having fixed the gauge by our choice of $U$ in Eq.~\eqref{Eq:SpinRotation}, we can consistently discuss the SAT contributions one by one below.

\subsection{Emergent spatial confinement to effective wires}\label{SubSec:Confinement}
The effective chemical potential after the transformation reads
\begin{equation}\label{Eq:MuEff}
\mu_\text{eff} = \mu - \frac{\inpar{\hbar g^\prime}^2}{8m} - \frac{\hbar}{2}\alpha g^\prime \sin f \,.
\end{equation}
Recall that $\sin f$ is either zero or one, such that the last term is only present in the cycloidal case. The renormalization of the chemical potential has been derived earlier for one-dimensional spin spirals \cite{KWF12} and skyrmions with sinusoidal radial profile \cite{YSK16}. Our system differs from these examples in that the correction to $\mu$ is not constant. Namely, the derivative $g^\prime=\partial g / \partial y$ varies periodically across the system whenever $B\neq 0$.

We define the ratio $\beta=B/|J\magn|$ of external to intrinsic exchange terms. In practice, we are mostly interested in cases where $0\leq\beta<1$, as the net field would only tumble rather than rotate for $\beta>1$ and the magnetic texture could no longer be expected to have a decisive effect. Exactly at $\beta=1$, $\Magn=0$ at some places in the plane, such that the SAT would become ill-defined. We obtain from Eq.~\eqref{Eq:gAngle}
\begin{equation}\label{Eq:GPrime}
g^\prime = \frac{2\pi}{l}\, \frac{\beta\cos(2\pi y/l)+1}{\inbra{\beta^2 + 2\beta\cos(2\pi y/l) + 1}}.
\end{equation}
Thus, $\mu_\text{eff}(y)$ oscillates between
\begin{equation}
\mu-\frac{\hbar^2}{8m}\inbra{\frac{2\pi}{l(1-\beta)}}^2
\leq \mu_\text{eff}(y) \leq
\mu-\frac{\hbar^2}{8m}\inbra{\frac{2\pi}{l(1+\beta)}}^2\,,
\end{equation}
skipping the last term of Eq.~\eqref{Eq:MuEff} for simplicity. Formally, the bound diverges as $\beta$ approaches one. On the other hand, when $g^\prime\rightarrow\infty$, our assumption of a slowly varying field breaks down. Still, one can expect a pronounced effect of the SAT corrections to $\mu$.

It is readily clear that $\mu_\text{eff}$ will enter the condition for band inversions and thereby topological phases, which we discuss later. However, even independently of topology, a large value of $\mu_\text{eff}$ at some $y$ will effectively act like a confinement potential for any low-lying states to stripes of low $\mu_\text{eff}$. Furthermore, if these stripes are sufficiently narrow, they will turn into quasi-one-dimensional wires --- a first prerequisite for the creation of Majorana wires with localized end modes. The basic principle of this confinement mechanism remains valid in arbitrary two-dimensional textures $\Magn(\pos)$ where effective topological wires have been discussed, originating from arrays of nanomagnets, skyrmions, or other structures \cite{FMS16, MSK17, ZMH19, GSK18}.

\subsection{Combined spin-orbit coupling}\label{SubSec:SOC}
A major effect of the SAT is to reveal the synthetic spin-orbit coupling \cite{BJK10, KWF12} of the form
\begin{equation}
h^\text{so}_\text{syn} = \hbar\alpha^\prime k_y \inpar{\sigma_x\sin f - \sigma_y \cos f}\,,
\end{equation}
which stems from the kinetic term in the Hamiltonian. We have introduced the coefficient
\begin{equation}
\alpha^\prime = \frac{\hbar g^\prime}{2m}\,.
\end{equation}
In the special case $B=0$, $\alpha^\prime$ is a constant, and otherwise a function of $y$. It is worthwhile to take closer look at the overall effective spin-orbit coupling, because the spin-orbit splitting close to the Fermi surface is essential for the size of the effective gap in the superconducting hybrid system. We will see that the combination of Rashba and synthetic contributions can cause \textsl{weak spots} in the gap.

\begin{figure*}[t]
\includegraphics[width=\textwidth]{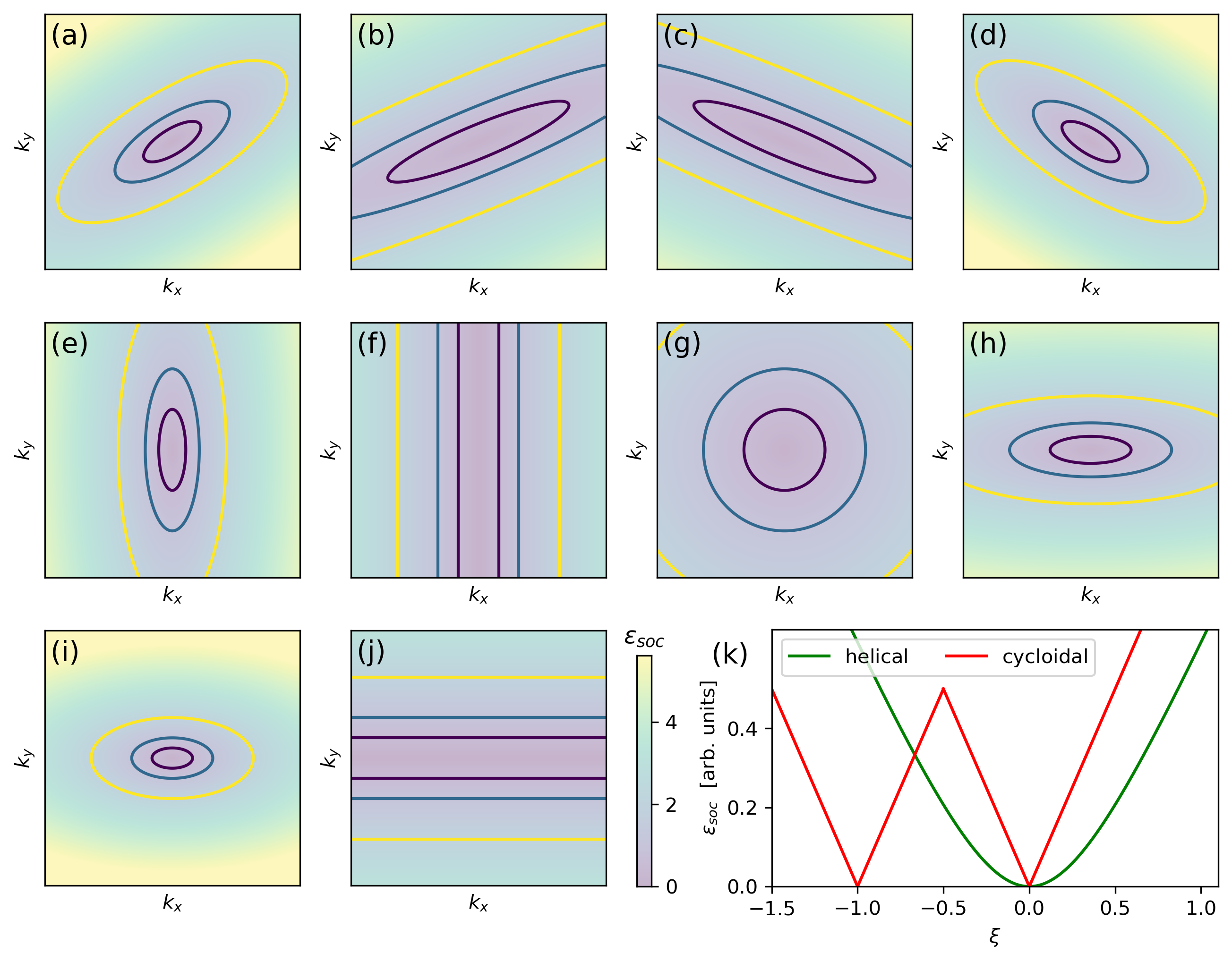}
\caption{\label{Fig:SOC} The combined effect of Rashba and magnetically induced SOC: momentum-space profile of the resulting energy splitting $\epsilon_\text{soc}=\hbar|\mathbf{s}(\mom)|$, cf. Eq.~\eqref{Eq:SOC-vector}, displayed by a continuous color scale and additionally by contour lines at the levels $\frac12$ (purple), $1$ (blue), and $2$ (yellow) in arbitrary units. [(a)--(d)] helical texture with $\xi=-1$, $-\frac12$, $\frac12$, and $1$, respectively; [(e)--(i)] cycloidal texture with $\xi=-\frac32$, $-1$, $-\frac12$, $\frac12$, and $1$, respectively; (j) $\xi=0$ for either texture; and (k) the minimal spin-orbit splitting energy on a circular Fermi surface for helical and cycloidal magnets as a function of $\xi=\alpha/\alpha^\prime$, the ratio of Rashba to synthetic spin-orbit coefficients.}
\end{figure*}

The momentum-dependent vector $\mathbf{s}(\mom)$ in Eq.~\eqref{Eq:HcontSAT} is either
\begin{equation}\label{Eq:SBloch}
\mathbf{s}_\text{heli}(\mom) =
\begin{pmatrix}
\alpha k_y \cos g  \\ -\alpha k_x -\alpha^\prime k_y \\ \alpha k_y \sin g
\end{pmatrix}
\end{equation}
for a helical magnet, or
\begin{equation}\label{Eq:SNeel}
\mathbf{s}_\text{cyc}(\mom) =
\begin{pmatrix}
\inpar{\alpha+\alpha^\prime}k_y  \\ -\alpha k_x \cos g \\ -\alpha k_x \sin g
\end{pmatrix}
\end{equation}
for a cycloidal magnet. The spin-orbit energy splitting $\epsilon_{soc} = \hbar|\mathbf{s}|$ depends only on the magnitude of $\mathbf{s}$, for which we find
\begin{equation}\label{Eq:SOC-vector}
|\mathbf{s}(\mom)|^2 = \inpar{\alpha^\prime}^2
\begin{cases}
 \xi^2 k_y^2 + \inpar{\xi k_x + k_y}^2 & \text{helical} \\
 \xi^2 k_x^2 + \inpar{\xi + 1}^2k_y^2  & \text{cycloidal,}
\end{cases}
\end{equation}
with the ratio $\xi=\alpha/\alpha^\prime$. In polar momentum coordinates $(k,\varphi)$,
\begin{equation}
|\mathbf{s}(k,\varphi)|^2 = \inpar{\alpha^\prime}^2 k^2
\begin{cases}
 \xi^2 + \xi\sin(2\varphi) + \frac12[1-\cos(2\varphi)]\\
 \xi^2 + \frac12[1-\cos(2\varphi)]\inpar{2\xi+1}.       
\end{cases}
\end{equation}
Of course, $|\mathbf{s}_\text{heli}| = |\mathbf{s}_\text{cyc}|$ in the absence of Rashba SOC, i.e., $\xi=0$.

Now consider a circular Fermi surface at $k=k_F$. The spin-orbit splitting at $k_F$ has minima at certain angles in momentum space, namely at
\begin{equation}
\varphi_\text{min, heli} = \pm\frac12 \text{arctan}\inpar{-2\xi}
\end{equation}
or at
\begin{equation}
\varphi_\text{min, cyc} =
\begin{cases}
\frac{\pi}{2} & \text{if } \xi < -\frac12 \\
0             & \text{if } \xi >  \frac12.
\end{cases}
\end{equation}
These angles define a weak axis of the effect, where the Rashba SOC counteracts the synthetic coupling induced from the winding of the magnetization. The minimal spin-orbit splitting at $\varphi_\text{min}$ is
\begin{equation}
\epsilon_\text{min, heli} = \hbar|\alpha^\prime|k_F
\sqrt{\xi^2 + \frac12\inbra{1 - \sqrt{1+4\xi^2}}}
\end{equation}
or
\begin{equation}
\epsilon_\text{min, cyc} = \hbar|\alpha^\prime|k_F
\begin{cases}
|\xi| & \text{if } \xi>-\frac12 \\
|\xi+1| & \text{else.}
\end{cases}
\end{equation}

The profile of the overall spin-orbit coupling in momentum space is shown in Figs.~\ref{Fig:SOC}(a)--(j) for various situations. The weak axis lies along the $x$ direction when $\xi=0$. Otherwise, it gets gradually tilted in the helical case, whereas it flips to the $y$ direction in the cycloidal case when $\xi$ becomes smaller than $-\frac12$. Exactly at this special point, the effect is isotropic. The sign of $\xi$ depends in our notation on the sign of $\alpha$. Alternatively, a sign change could be understood as an inversion of the magnetic winding direction which can be invoked through $f \rightarrow f+\pi$ (which makes no difference in the helical case).

The minimum of the spin-orbit splitting is shown in Fig.~\ref{Fig:SOC}(k). A simple calculation shows that $\epsilon_\text{min, heli} \leq \epsilon_\text{min, cyc}$ whenever $\xi>-\frac23$. In particular, in the case of weak Rashba coupling, cycloidal winding provides a much larger minimal spin-orbit splitting on the Fermi surface than helical winding.

\subsection{Gap closings and topological regions}\label{SubSec:GapClosings}
Now we return to the full transformed Hamiltonian in Eq.~\eqref{Eq:HcontSAT}. Exact analytic expressions for the eigenenergies exist, but are not insightful in the general case. It is possible, though, to carve out useful information about the existence and location in $\mom$ space of bulk zero-energy states. Such states may appear, e.g., in nodal phases or when the gap closes and re-opens in the event of a topological transition.

The eigenenergies follow from the characteristic polynomial, which is reduced to the bare determinant of $\hat{\tilde H}$ for zero-energy solutions. The task is then to evaluate whether a real-valued momentum $\mom_0$ exists such that $\text{det}\inbra{\hat{\tilde H}(\mom_0)}=0$, where
\begin{eqnarray}
\text{det}\inbra{\hat{\tilde H}(\mom)} &=& 
\inbra{\inpar{\frac{\hbar^2k^2}{2m} -\mu_\text{eff}}^2 + \Delta^2 - M^2 - \hbar^2\mathbf{s}(\mom)^2}^2 \notag\\
&&{} + 4\hbar^2\inbra{\Delta^2\mathbf{s}(\mom)^2 - M^2s_z(\mom)^2}, \label{Eq:Det}
\end{eqnarray}
and the spin-orbit vector $\mathbf{s}$ is given by Eqs.~\eqref{Eq:SBloch} and \eqref{Eq:SNeel}.

Let us first recover the criterion for band inversion at zero momentum, which is well known for Majorana nanowires \cite{LSD10, ORO10, KWF12} as well as two-dimensional topological hybrid systems. At $\mom=0$, the kinetic and spin-orbit terms vanish, such that $\text{det}\inbra{\hat{\tilde H}(\mom=0)}=\inbra{\mu_\text{eff}^2 + \Delta^2 - M^2}^2$. Consequently, the gap closes where
\begin{equation}\label{Eq:InvAtZero}
M(\pos)^2 = \Delta^2 + \mu_\text{eff}(\pos)^2 \,,
\end{equation}
with a band-inverted state for larger $M$. Together with Eq.~\eqref{Eq:MuEff}, this amounts to a criterion to identify contours of \emph{topological regions} in two dimensions, which has been applied in several recent articles \cite{FMS16, MSK17, ZMH19, GSK18} for other textures. Band inversion is facilitated if $\mu$ is chosen such that $\mu_\text{eff}$ is close to zero in the topological regions.

\begin{figure*}[t]
\includegraphics[width=\textwidth]{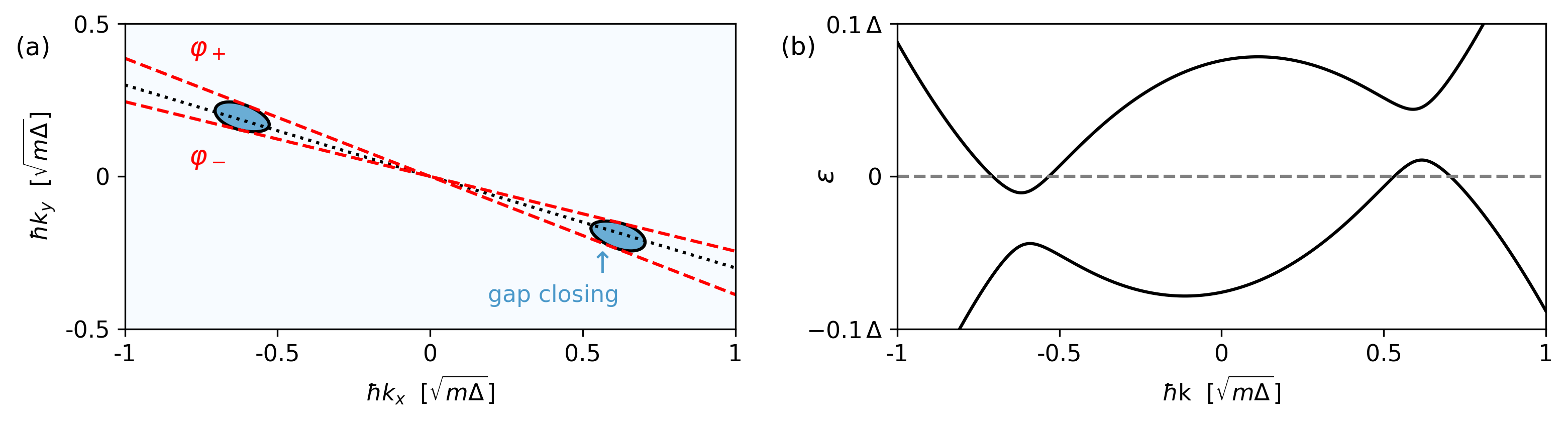}
\caption{\label{Fig:indirect}Indirect gap closing: (a) In momentum space, the regions where the gap closes (blue shaded) occur for angles $\varphi\in[\varphi_-,\varphi_+]$ (see main text) as indicated by the red dashed lines. (b) The two low-energy levels of the transformed BdG Hamiltonian, Eq.~\eqref{Eq:HcontSAT}, for momenta along a cut through the gap closing regions [black dotted line in panel (a)]. In this plot, $|J\magn|=1.25\,\Delta$, $B=0.18\,\Delta$, $\mu=0$, $y=l/4$, and $\xi=0.3$.}
\end{figure*}

Now we discuss gap closings at $\mom \neq 0$. We consider first cases where $\mathbf{s}(\mom)=0$ at the angle $\varphi_\text{min}$, as discussed in the previous subsection. This happens either if $\alpha=0$ (then $\varphi_\text{min}=0$) or if the texture is cycloidal and $\xi=-1$ (then $\varphi_\text{min}=\frac{\pi}{2}$). Then the second line in Eq.~\eqref{Eq:Det} vanishes, and the first line can be solved exactly, yielding the momenta
\begin{equation}\label{Eq:ZeroSolutions}
k_{1,2,3,4} = \pm\frac{1}{\hbar}\sqrt{2m\inpar{\mu_\text{eff}\pm\sqrt{M^2-\Delta^2}}}
\end{equation}
at which the gap vanishes. The two signs $\pm$ are independent of each other. The number of \emph{real} solutions on the weak axis of the net SOC (appearing as symmetric pairs) depends on $|M|$:
\begin{enumerate}
\item If $M^2 < \Delta^2$, then no solution exists.
\item If $\Delta^2 < M^2 < \Delta^2 + \mu_\text{eff}^2$, there are four solutions if $\mu_\text{eff}>0$ and none if $\mu_\text{eff}<0$.
\item If $M^2>\Delta^2+\mu_\text{eff}^2$, there are two solutions.
\end{enumerate}
Equalities in these expressions cause degenerate solutions. The first case is continuously connected to the trivial superconductor via $M\rightarrow 0$. The second case has been observed numerically \cite{SAB15}, but does not involve band inversion at $\mom=0$ and is therefore not covered by the notion of topological regions (if $\mu_\text{eff}>0$). The third case, in contrast, is related to Eq.~\eqref{Eq:InvAtZero}. At the transition point, Eq.~\eqref{Eq:ZeroSolutions} yields $k=0$, and this solution splits into two nodes when $M^2$ trespasses $\Delta^2+\mu_\text{eff}^2$.

In both the second and the third cases, the gap closings represent persisting nodal points in the effective superconducting gap. Reversely, a continual topological gap strictly demands nonzero total SOC even at $\varphi_\text{min}$ and thereby nonzero Rashba coupling. We note that there are different magnetic textures, like skyrmions \cite{YSK16,GSK18,RGM19,GMS19,MBG20}, where Rashba coupling is not necessary for a fully gapped phase. The topological phases with point nodes exhibit flat bands of Majorana edge modes, as demonstrated in Ref.~\cite{SAB15} for the case $B=0$ and $\alpha=0$. On an arbitrarily oriented edge, the intervals of edge momenta where flat bands exist are bounded by the projections of the bulk nodal points.

Finally, we turn to the generic situation where nonzero total SOC is found in any momentum space direction. In a band-inverted state, the full square in the determinant [first line of Eq.~\eqref{Eq:Det}] will still become zero at some $\tilde{k}>0$. The existence of bulk zero-energy states is then entirely determined by the sign of the expression in the second line of Eq.~\eqref{Eq:Det}, similar to the discussion in Ref.~\cite{ReS14}. The term $\propto\Delta^2\mathbf{s}^2>0$ amounts to the spin-orbit-assisted gap opening by $s$-wave pairing but is counteracted by the term $\propto -M^2 s_z^2$. Interestingly, the latter term involves only the Rashba contribution to the spin-orbit coupling.

The sign depends only on $\varphi$ and not on $k$. It is straightforward to show that the sign changes at angles in momentum space where
\begin{equation}\label{Eq:WeakSpot}
\tan\varphi_\pm =
\begin{cases}
\xi\inbra{-1\pm\xi\sqrt{\frac{M^2\sin^2 g}{\Delta^2}-1}}^{-1} & \text{helical} \\
\frac{\pm\xi}{\xi+1}\sqrt{\frac{M^2\sin^2 g}{\Delta^2}-1} & \text{cycloidal.}
\end{cases}
\end{equation}
As the radicand becomes positive, an interval $[\varphi_+,\varphi_-]$ opens up in which extended zero-energy bulk states exist for momenta close to $\tilde{k}$. This is illustrated in Fig.~\ref{Fig:indirect}. If a very large amount of bulk bands were to be included in the model, one could even expect a finite area in momentum space filled with zero-energy bulk states. We note that this is an indirect closing of the gap which does not involve band crossings.

In conclusion, the system becomes gapless in regions where the in-plane component of $\Magn$ can trespass the gap, i.e., $M^2\sin^2 g > \Delta^2$, while band-inversion appears simultaneously. This result is reminiscent of the indirect gap-closing condition in the one-dimensional case of Majorana nanowires in a tilted magnetic field \cite{ReS14, ORS14}, where it has been confirmed experimentally \cite{GZB18}. Given that the external field $B$ does not contribute to the in-plane component of $\Magn$, we obtain the simple relation
\begin{equation}
|J\magn| < |\Delta|,
\end{equation}
which ensures a full bulk gap in topological regions of the system.

In the absence of gap closings, Eq.~\eqref{Eq:WeakSpot} still provides information about where to expect weak spots of the gap. In the helical case, this will be at $\varphi\approx-\arctan\xi$ and in the cycloidal case on the $x$ axis, for some momentum $k\approx k_F$. For small $\xi$, these weak spots coincide with the weak axis of spin-orbit splitting up to a difference in $\varphi$ of order $\xi^3$. For large $\xi$, this is not the case. In fact, we see that the Rashba term plays an ambivalent role: It is required to prevent the total spin-orbit splitting from being zero along one axis and thereby ensures a full gap. At the same time, though, it weakens the gap at other momenta. In consequence, the effective topological gap is typically much smaller than the original $s$-wave gap.

\subsection{Effective Majorana wires: conditions}
Let us briefly summarize the continuum conditions for emergent Majorana nanowires that we have identified so far for helical or cycloidal magnetic textures:
\begin{enumerate}
\item Ensure emergent confinement to stripes by tuning of the external field with respect to $|J\magn|$.
\item Confined regions become topological by band-inversion at $k=0$ where $\Magn(y)^2 > \Delta^2 + \mu_\text{eff}(y)^2$.
\item Rashba SOC is required in addition to the synthetic SOC in order to avoid nodal points in the gap.
\item $|J\magn|<\Delta$ guaranties that the effective gap cannot close indirectly.
\item Confined and topological stripes must be sufficiently narrow to form effective wires.
\end{enumerate}

\begin{figure}[t]
\includegraphics[width=\columnwidth]{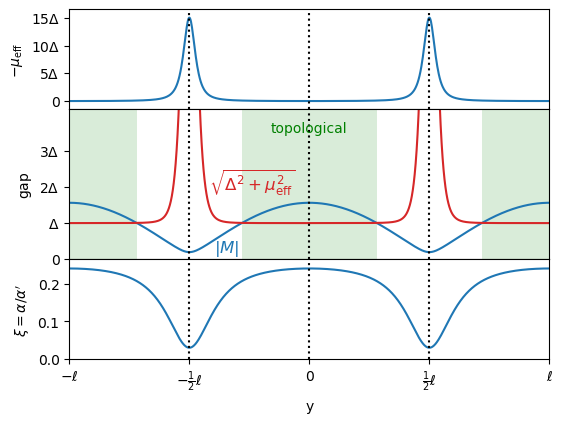}
\caption{\label{Fig:Regions}A case in favor of emergent Majorana wires within the continuum model. The negative $\mu_\text{eff}$ (top panel) creates potential barriers which slice the system into wires. The magnitude of the total exchange field satisfies the band inversion condition (middle panel) in topological regions (green) within the wires. This is facilitated by choosing $\mu$ such that $\mu_\text{eff}\approx0$ inside the wires. The lower panel shows the oscillations of the ratio of spin-orbit couplings.}
\end{figure}

\begin{figure}
\includegraphics[width=\columnwidth]{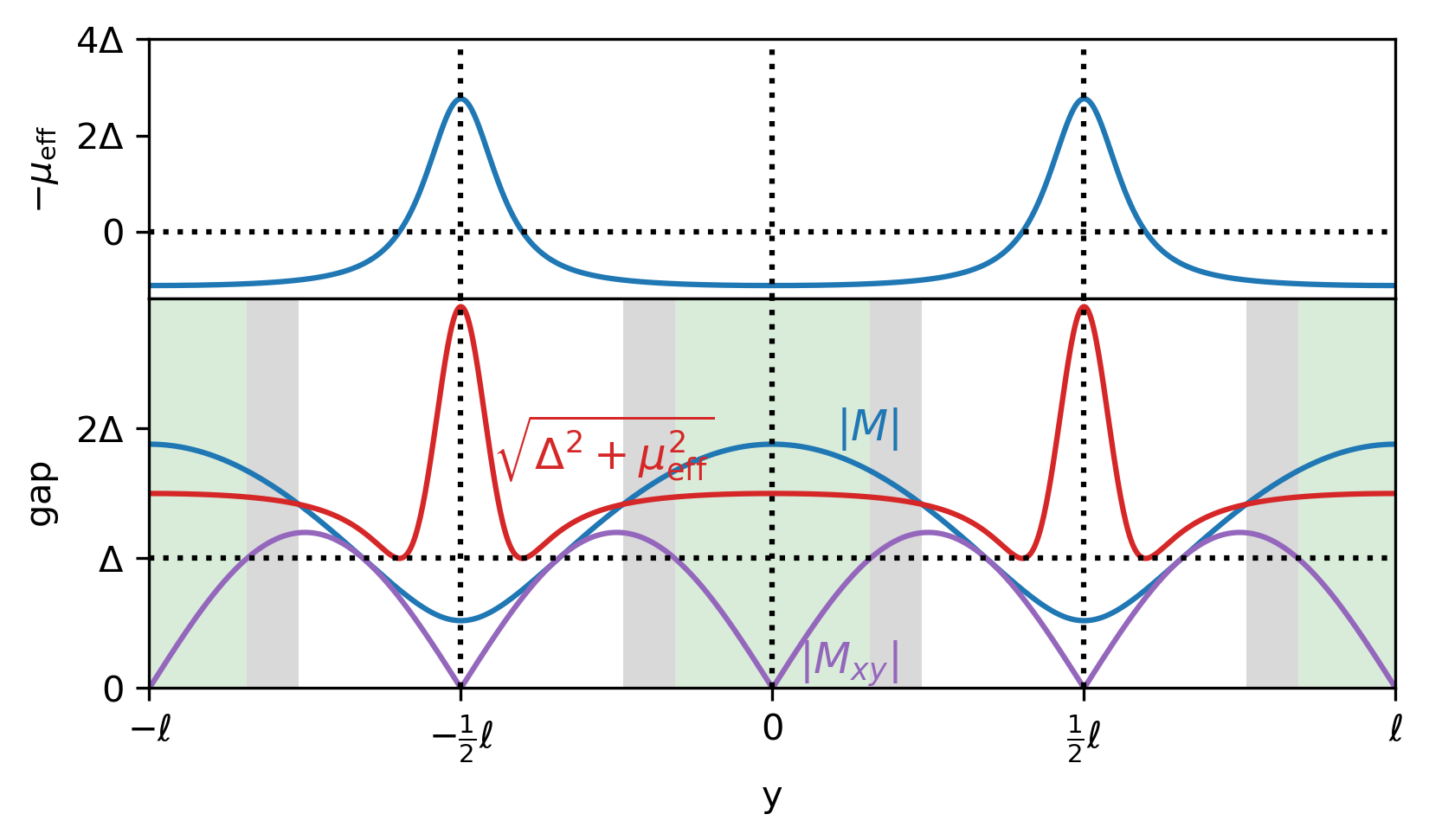}
\caption{\label{Fig:BadRegions}A case with unfavorable parameters to observe Majorana modes. In particular, $J|\magn|>\Delta$ and the in-plane component of the exchange field  $M_{xy}$ trespasses $\Delta$ in in some part (gray shaded) of the topological region, leading to an indirect band gap closure. Thus, although band inversion still takes place, zero-energy states would hybridize with bulk modes.}
\end{figure}

The relevant position-dependent quantities are depicted in Fig.~\ref{Fig:Regions} for a parameter choice in favor of Majorana wires, whereas Fig.~\ref{Fig:BadRegions} shows an unfavorable situation. We suggest that the notion of topological contours should not be used independently of the other criteria in search of possible Majorana modes. In particular, in previous work using such contours, they likely coincided approximately with the confinement barriers and could therefore be consistently be interpreted as effective wires.
\begin{figure}[t]
\includegraphics[width=0.75\columnwidth]{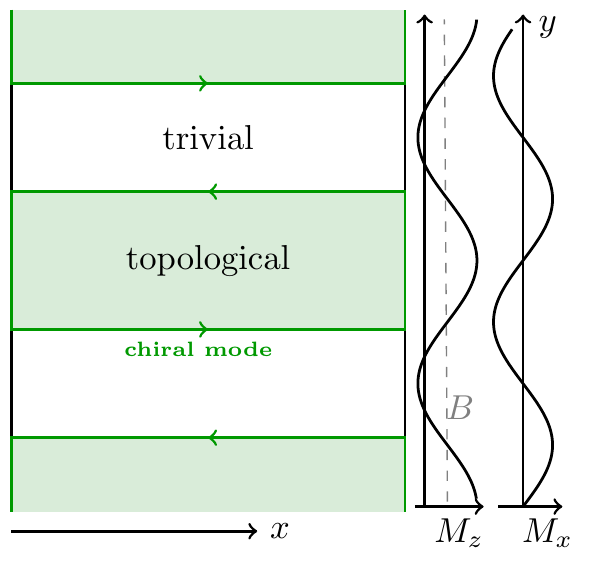}
\caption{\label{Fig:TopView}Top view of the system for the case shown in Fig.~\ref{Fig:Regions}. The regions where topological band inversion takes place are highlighted in green. Without hybridization, one would expect one-dimensional chiral modes between topological and trivial regions (green arrows).}
\end{figure}
A remark is at hand concerning the last condition, which was only mentioned briefly in Sec.~\ref{SubSec:Confinement}: The confined and topological stripes appear as two-dimensional regions, rather than wires, within the present model. Thus, in the first place we expect chiral Majorana modes propagating along the edges of these regions, as shown in Fig.~\ref{Fig:TopView}. However, a crossover to a quasi-one-dimensional regime with localized MBS at the ends of a stripe is possible \cite{MCR19}. The chiral zero-energy modes are exponentially localized at the edge with a localization length which is inversely proportional to the effective gap, $\xi_\text{loc}\approx\hbar v_F/\Delta_\text{eff}$. If the width of a topological stripe is small compared to $\xi_\text{loc}$, chiral modes at opposite edges hybridize and leave only two unpaired MBS at the ends of the stripe. We expect that this condition is always satisfied in reality. For example, a gap of roughly $1\,\text{K}$ would lead to a localization length on the order of $10\,\mu\text{m}$, whereas the magnetic period $l$ is typically on the order of $10\hdots100\,\text{nm}$ \cite{HDL18,SMK18,KBW20}. Practically, the challenge might reversely be to maximize the effective size of the topological gap in order to avoid a significant overlap of the end states of emergent wires.

In the next section, we drop the approximation of a slowly varying field and move on to a tight-binding description in terms of the magnetic unit cell --- at the expense that the precise conditions on the parameters are no longer transparent.

\section{$C_2$-symmetric tight-binding model}\label{Sec:C2TB}
With the continuum model, we have demonstrated that helical or cycloidal magnet-superconductor hybrids can host a set of emergent parallel Majorana nanowires if the system parameters are chosen suitably. Now we connect this idea to the theory of crystalline topological phases \cite{Fu11, CTS16}. In our case, the \quotmarks{crystalline} symmetry is provided by the periodic magnetic texture, though. More precisely, we set up a tight-binding model where the unit cell spans over one period of magnetic rotation in the $y$ direction. The Hamiltonian has $C_2$ symmetry and can be classified by the respective set of topological crystalline indices.

\subsection{Hamiltonian and parameter relations}
For computational ease, we assume an atomic square lattice with lattice constant $a$. Furthermore, we set up the Hamiltonian assuming that the magnetic and the atomic lattices are aligned and commensurate, such that one unit cell of size $a\unitvec_x\times l\unitvec_y$ consists of $n=l/a$ internal sites. We place the unit cell in the $y$ direction such that the magnetization vector and the background field are parallel at the edges and antiparallel at the center.

The Hamiltonian on the Brillouin zone of size $\frac{2\pi}{a}\inpar{1 \times \frac{1}{n}}$ is then represented by a matrix in a space of dimension $2_\sigma\times2_\tau\times n$, accounting for the spin, Nambu, and internal site dimensions. At each internal site, we work in the same spinor basis as in the continuum model, $\inpar{c_{\mom,\uparrow},c_{\mom,\downarrow},c_{-\mom,\uparrow}^\dagger,c_{\mom,\downarrow}^\dagger}$. The contributions to the Hamiltonian are the discretized version of the terms appearing in Eq.~\eqref{Eq:BdGRealSpace} and are implemented without prior SAT. Hopping is restricted to nearest neighbors. All terms are listed explicitly in Appendix~\ref{App:TB}.

As a helping hand to connect the tight-binding model and the continuum model, we summarize all parameters and their conversion in Table~\ref{Tab:Param}. To avoid ambiguity, and for ease of notation, we will only use $\mu=\mu_\text{cont.}$ in the main text. With this, we can also express the two effective parameters $\mu_\text{eff}$ and $\xi$ by tight-binding quantities. Both depend on the real-space position via $g^\prime(y)$ in the continuum model. Averaging over $y$, i.e., $g^\prime=2\pi/l$, we get from Eqs.~\eqref{Eq:MuEff} and \eqref{Eq:GPrime}
\begin{equation}\label{Eq:MuEffTB}
\mu_\text{eff} = \mu - \frac{\pi^2 t}{n^2}
                 - \frac{\pi\lambda}{n}\sin f\,.
\end{equation}
As mentioned earlier, expect that a good choice of $\mu$ is to compensate approximately for the correction terms, such that $\mu_\text{eff}\approx 0$, which minimizes the Zeeman splitting required for band inversion at $k=0$.

The averaged ratio of the intrinsic to the induced spin-orbit coupling reads
\begin{equation}\label{Eq:XiTB}
\xi = \frac{n\lambda}{2\pi t} \,,
\end{equation}
which allows one to compare the overall SOC qualitatively to the cases shown in Fig.~\ref{Fig:SOC}.

\begin{table}
\caption{\label{Tab:Param}Synopsis of parameters of the continuum and the tight-binding model and their relations assuming an atomic square lattice.}
\begin{ruledtabular}
\begin{tabular}{lll}
Continuum      & Tight-binding          & Relation \\\hline
---            & Lattice const. $a$     & $a_\text{cont.}\longrightarrow 0$ \\
Mass $m$       & Hopping $t$            & $m = \hbar^2/(2ta^2)$ \\
Chem. pot. $\mu_\text{cont.}$ & Chem. pot. $\mu_\text{t.b.}$  & $\mu_\text{cont.}\!= \mu_\text{t.b.}\!+4|t|$ \\
Period length $l$ & Atoms per cell $n$  & $l=na$ \\
Position $y$   & --- & Only cont. \\
Rashba coeff. $\alpha$ & Rashba const. $\lambda$ & $\alpha = a\lambda/\hbar$ \\
\multicolumn{2}{c}{Exchange coupling $J$}     & Identical \\
\multicolumn{2}{c}{External field $B$}        & Identical \\
\multicolumn{2}{c}{$s$-wave pairing $\Delta$} & Identical \\
\multicolumn{2}{c}{Helical/cycloidal $f\in\{0,\frac12\pi\}$} & Identical
\end{tabular}
\end{ruledtabular}
\end{table}

\subsection{Symmetries and topological invariants}
We will now identify the relevant symmetry-protected topological invariants of the system. The symmetries discussed below are also present in the continuum model. However, within the slowly varying field approximation, we could have defined only local invariants. In contrast, the invariants of the tight-binding model define global topological phases.

The BdG Hamiltonian has particle-hole symmetry by construction,
\begin{equation}
\Xi H_\text{t.b.}(\mom) \Xi^\dagger = -H_\text{t.b.}(-\mom)\,,
\end{equation}
with the antiunitary operator $\Xi=t_x\mathcal{K}$. Time-reversal symmetry is broken by the net magnetic field. Thus, the system belongs to class $D$ \cite{AlZ97, EvM08} and has a (strong) topological $\mathbb{Z}$ invariant in two dimensions \cite{RSF10, CTS16}, namely, the Chern number
\begin{equation}
\chern = \frac{i}{2\pi}\int_\text{BZ} \text{Tr}_{\varepsilon<0}\inpar{d\mathcal{A}+\mathcal{A}\wedge\mathcal{A}} \in \mathbb{Z}\,,
\end{equation}
with the Berry connection $\mathcal{A}^{\alpha\beta}=\inang{u^\alpha(\mom)|d u^\beta(\mom)}$. Numerically, we evaluate first $\chern$ per band by adding up plaquette Berry fluxes in the Brillouin zone and sum subsequently over occupied ($\varepsilon<0$) states \cite{AOP15}.

In addition to $\Xi$, we can identify a unitary symmetry. Helical or cycloidal magnetic order is consistent with a twofold rotation ($C_2$) symmetry. It is mediated by the operators $\hat{R}=-\unity$ acting on vectors in the $xy$ plane and
\begin{equation}
\hat{r} = e^{i\frac{\pi}{2}\sigma_z}\tau_z(-\mathbb{A}) = -i\sigma_z\tau_z\mathbb{A}
\end{equation}
acting on the internal degrees of freedom, where $\mathbb{A}$ is the antidiagonal in $n\times n$ space of sites per unit cell. It is easy to check that $\Xi$ and $\hat{r}$ commute, and that
\begin{equation}\label{Eq:C2symm}
\hat{r}^\dagger H_\text{t.b.}(\hat{R}\mom) \hat{r} = H_\text{t.b.}(\mom)
\end{equation}
as long as two conditions are satisfied: (i) the external field $\mathbf{B}$ does not have an in-plane component, which we have therefore excluded from the beginning, and (ii) $\hat{r}$ is implemented such that $\Magn$ points in the out-of-plane direction at the rotation center. In our convention, the rotation center coincides with the center of the unit cell.

The presence of a unitary symmetry allows us to apply the framework of topological crystalline phases \cite{Fu11, CTS16}, under the condition that the bulk is fully gapped. There are four inequivalent rotation-invariant points $K\in\{\Gamma, X, M, Y\}$ in the Brillouin zone, where $\hat{R}K=K$. These are $\Gamma=(0,0)$, $X=(\pi,0)$, $M=(\pi,\pi/n)$, and $Y=(0,\pi/n)$. At these momenta, $H_\text{t.b.}(\mom=K)$ commutes with $\hat{r}$ according to Eq.~\eqref{Eq:C2symm}. Consequently, for eigenstates at $K$, their $\hat{r}$ eigenvalue is a good quantum number. This eigenvalue can only be $\pm i$ because $\hat{r}^2=-1$. Now, we denote the number of $+i$ states below zero energy by $n_K$. Then, the three indices
\begin{eqnarray}
\inbra{X} &=& n_X - n_\Gamma \\
\inbra{M} &=& n_M - n_\Gamma \\
\inbra{Y} &=& n_Y - n_\Gamma
\end{eqnarray}
are topological invariants of the system that are protected by the $C_2$ symmetry \cite{BTH14}. One should keep in mind that $\inbra{X}, \inbra{M}, \inbra{Z}$ are ambiguous in the sense that they depend on how the unit cell is chosen. What is actually invariant is the physical information about the existence of topological zero-energy modes, which is carried jointly by the set of all invariants (cf. Sec.~\ref{Sec:Disclinations}). 

It has been proven that any set of topological invariants $(\chern\,|\, [X], [M], [Y])$ has to obey the relation \cite{BTH14}
\begin{equation}\label{Eq:Restriction}
\chern + [X] + [Y] + [M] = 0 \mod 2\,.
\end{equation}
Hence, the crystalline indices determine whether even or odd Chern numbers are allowed.

\subsection{Topological phase diagram}
\begin{figure*}[pt]
\includegraphics[width=0.7\textwidth]{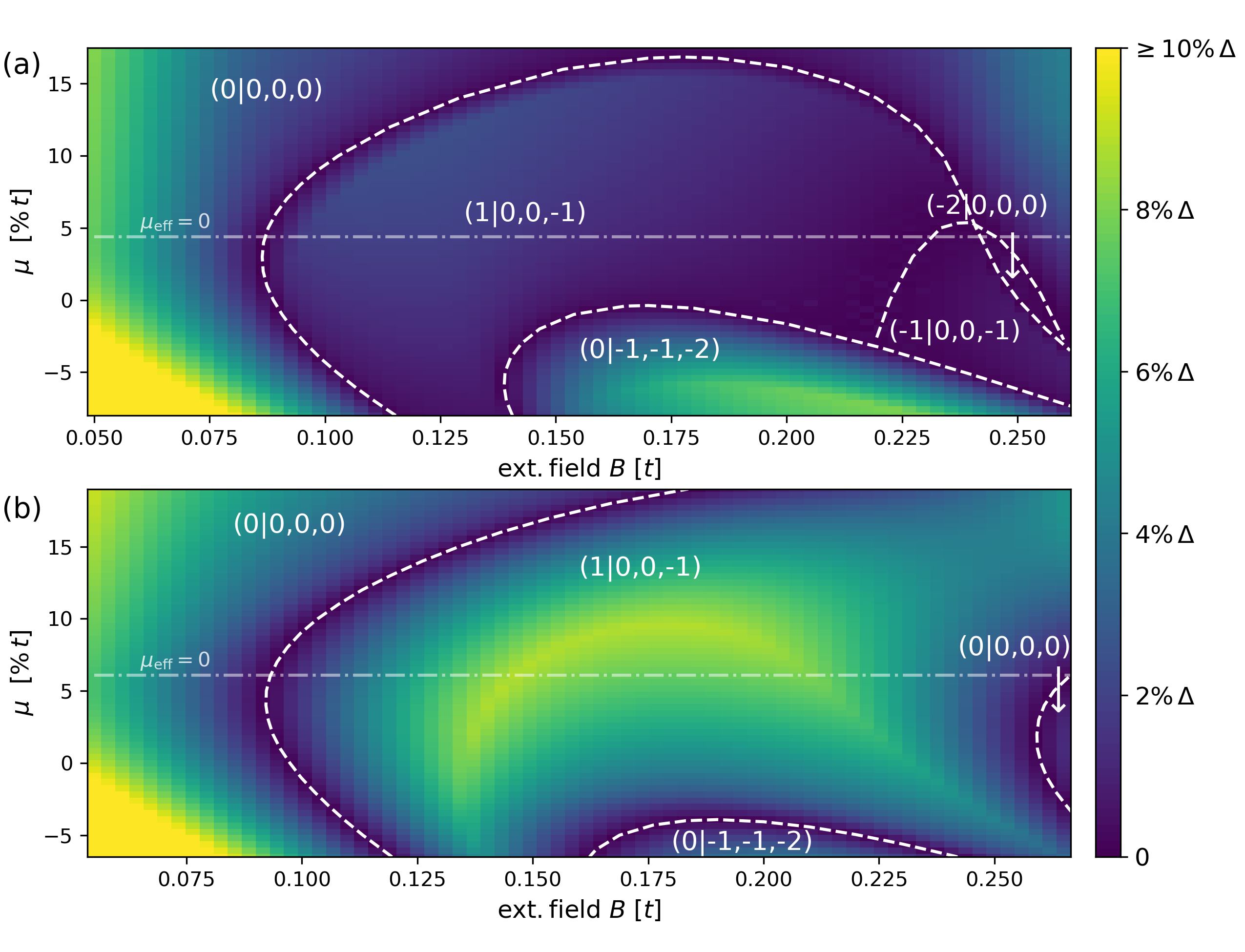}
\caption{\label{Fig:Phases-B-mu}The size of the effective gap compared to $\Delta$ as a function of the external field $B$ and the chemical potential $\mu$: (a) helical magnetization and (b) cycloidal magnetization. The white dashed lines indicate topological phase transitions (calculated independently). The topological phases are labeled by the tuple $(\chern\,|\,[X],[M],[Y])$. At the dash-dotted lines, $\mu_\text{eff}=0$. Here, $\Delta=0.25\,t$, $J=0.22\,t$, $n=15$, and $\lambda=0.085\,t$.}
\end{figure*}
\begin{figure*}[pt]
\includegraphics[width=0.7\textwidth]{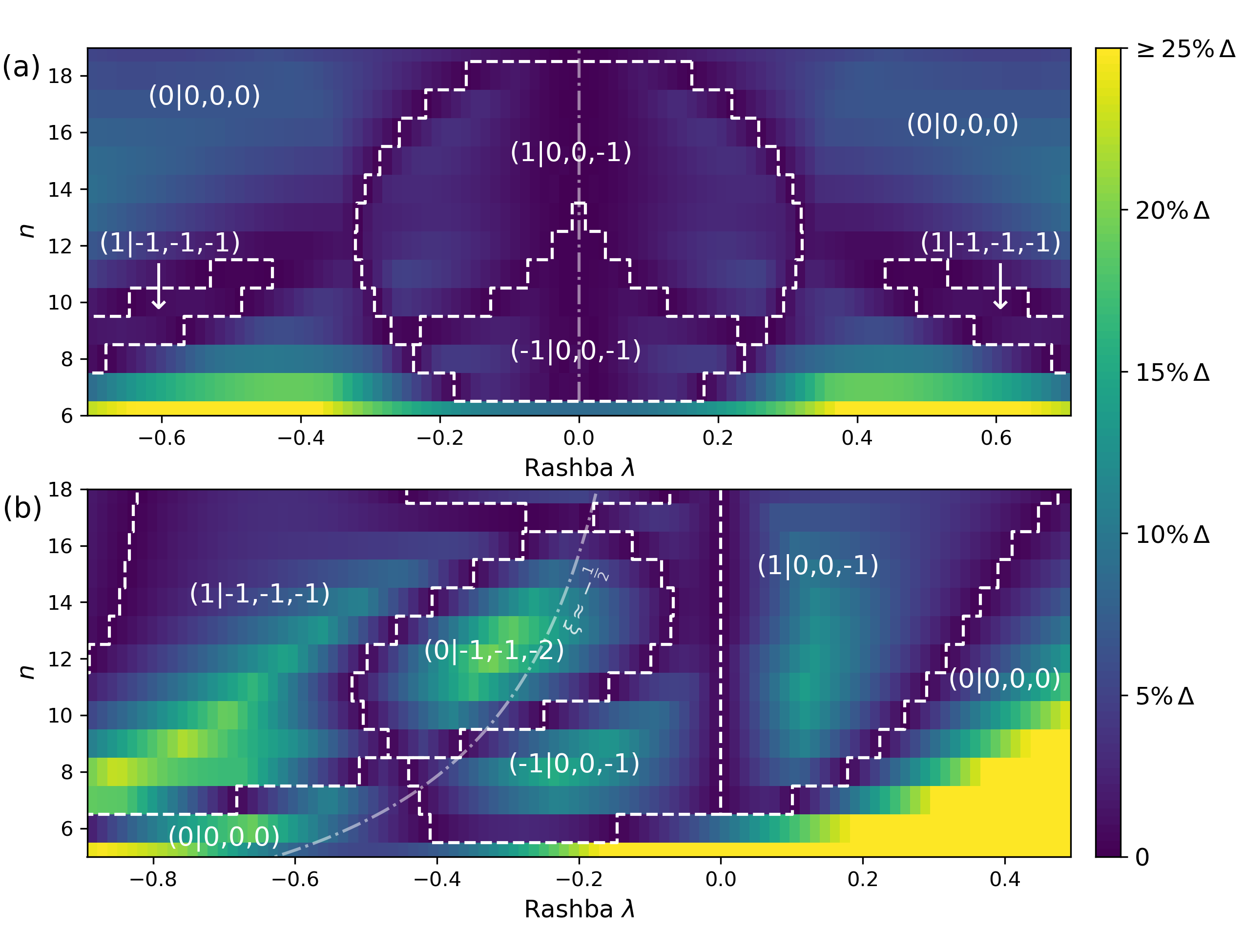}
\caption{\label{Fig:Phases-lambda-n}The size of the effective gap compared to $\Delta$ as a function of the Rashba constant $\lambda$ and the size of the magnetic unit cell $n$: (a) helical magnetization and (b) cycloidal magnetization. The white dashed lines indicate topological phase transitions (calculated independently). The topological phases are labeled by the tuple $(\chern\,|\,[X],[M],[Y])$. The dash-dotted line marks a gap closing without phase change at $\lambda=0$ in (a) and $\xi=-\frac12$ in (b). Here, $\Delta=0.25\,t$, $J=0.22\,t$, $B=0.17\,t$, and $\mu=0.04\,t$.}
\end{figure*}
\begin{figure*}[t]
\includegraphics[width=\textwidth]{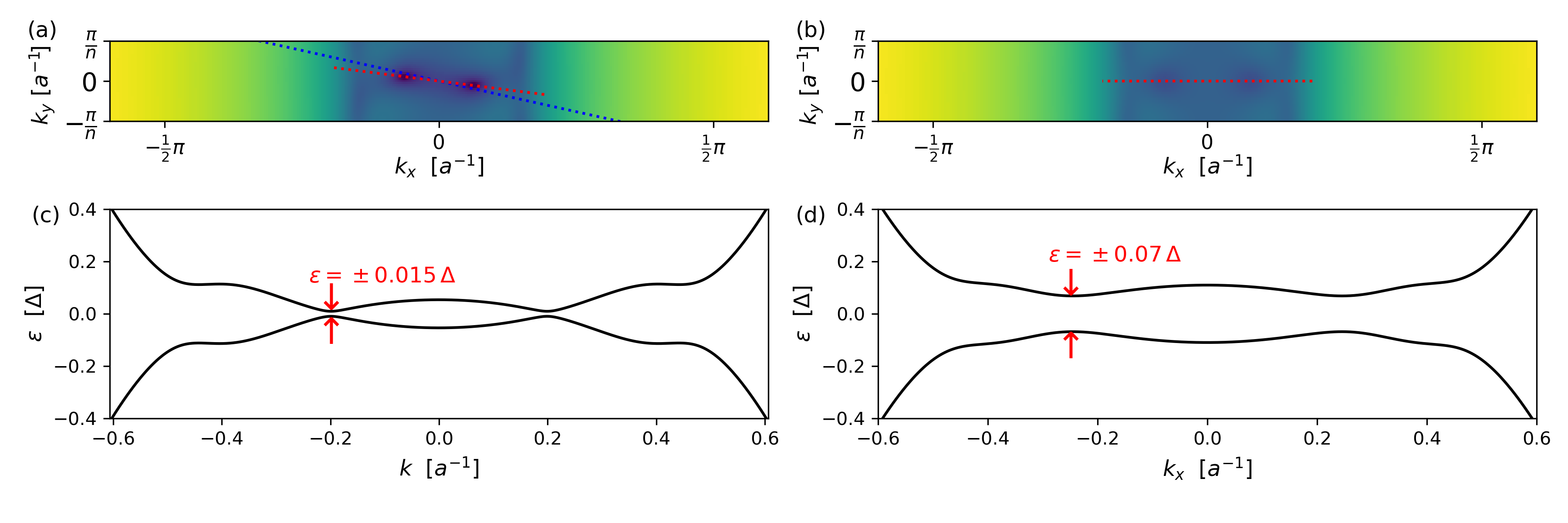}
\caption{\label{Fig:GapTB}The effective gap in momentum space. [(a) and (b)]: lowest eigenvalue, displayed by a logarithmic color scale, $\text{log}_{10}\frac{\varepsilon_\text{min}}{\Delta}$, for better visibility of the weak spots, for helical and cycloidal texture, respectively. Not the entire Brillouin zone is shown along $k_x$, but the gap is large elsewhere. In panels (c) and (d), the dispersion of the bands closest to zero energy are shown along a line in momentum space through the gap minima, as indicated by red dotted lines in panels (a) and (b), respectively. Additionally, in panel (a), the blue dotted line indicates the angle where the weak spots would be expected according to the continuum model, using the average value of $\xi$, cf. Eq.~\eqref{Eq:XiTB}. The parameters in this plot are $\Delta=0.25\,t$, $J=0.22\,t$, $B=0.17\,t$, $\mu=0.04\,t$, $\lambda=0.085\,t$, and $n=15$.}
\end{figure*} 
Now we study the topological phase diagram. In Fig.~\ref{Fig:Phases-B-mu}, the phases in the $B$-$\mu$ plane are shown together with the size of the effective gap. A similar plot of the $\lambda$-$n$ plane is shown in Fig.~\ref{Fig:Phases-lambda-n}. We find several phases with distinct topological indices. However, at least for the chosen parameters, most phases extend over a sizable region instead of forming a mosaic \cite{RoO15,RoO16,MBG20} of tiny (perhaps even fractal) phases. This is important for practical purposes, where the effective gap should not be too small and rough tuning of the system should be sufficient to reach the desired state.

For both helical and cycloidal magnetization, a strong topological phase with the signature $(1|0,0,-1)$ can be found. We will shortly argue that it corresponds to the heuristic picture of emergent Majorana wires from the continuum model. Therefore, we will focus on that phase. We find that the topological gap is generally larger in the cycloidal case compared to the helical case. This is consistent with our continuum discussion of the strength of the combined spin-orbit splitting. Therefore, cycloidal systems are favorable for our proposal. Theoretically, though, the same topological phase exists also in the helical case.

In general, the gap closings of the tight-binding Hamiltonian cannot be predicted from the continuum model, which does not account for the periodicity of the crystalline system. In particular, the band-inversion condition at the $\Gamma$ point alone, Eq.~\eqref{Eq:InvAtZero}, is insufficient to describe topological phases. For instance, in the $B$-$\mu$ plane, Fig.~\ref{Fig:Phases-B-mu}, the gap closes at the $Y$ point at the transition from the trivial to the $(1|0,0,-1)$ phase. At the transition to the $(0|-1,-1,-2)$ phase, it closes at the $\Gamma$ point. In the helical case, the gap closes along a further line, which starts approximately where $B$ exceeds $J$ (tumbling rather than rotating $\Magn$). Along this transition line, the gap closing points in momentum space lie on a tilted axis related to the weak spots of the gap. The gap remains very small in the $(-1|0,0,-1)$ and $(-2|0,0,0)$ phases, though, such that they may not be observable. The $(1|0,0,-1)$ phase reaches its maximal extent in the $B$ direction roughly when $\mu_\text{eff}=0$ according to Eq.~\eqref{Eq:MuEffTB}, in agreement with the expectation from the continuum model.

In the $\lambda$-$n$ plane, Fig.~\ref{Fig:Phases-lambda-n}, we find a symmetric diagram with respect to $\lambda$ in the helical case, whereas a pronounced asymmetry appears in the cycloidal case. This is consistent with our discussion of the overall spin-orbit coupling in Sec.~\ref{SubSec:SOC}; see, e.g., Fig.~\ref{Fig:SOC}(k). In fact, the phase diagram appears to be roughly symmetric around $\xi\approx-\frac12$ with respect to $\lambda$. This line, according to Eq.~\eqref{Eq:XiTB}, is also indicated in the figure. Given that cycloidal winding is preferable regarding the gap size, we emphasize that it is important in practice to be aware of the relative sign of the Rashba coupling and the synthetic SOC.

At $\lambda=0$, we obtain a gapless state (unless for very small $n$), which does, however, only coincide with a topological transition when the magnetization is cycloidal. In the helical case there are, again, distinct topological phases with a tiny gap. In the cycloidal case, negative and weak $\lambda$ leads to a phase with reversed Chern number, whereas very strong negative $\lambda$ allows for additional phases. Although one may expect large Rashba coupling to be favorable, it turns out that a very strong $\lambda$ (any sign) eventually triggers a transition to the trivial phase.

For parameters chosen within the $(1|0,0,-1)$ phase, Fig.~\ref{Fig:GapTB} shows the momentum-space structure of the effective gap for both helical and cycloidal magnetic textures. It is clearly visible that the gap has weak spots, as expected from our continuum analysis. The weak spots are more pronounced in the helical case compared to the cycloidal case, given the weaker overall spin-orbit splitting. The position of the gap minima is qualitatively in agreement with the continuum model. Quantitatively, though, the weak spots appear closer to the $x$ axis in the tight-binding calculation than predicted by the continuum model for the helical texture. For the cycloidal case, they lie exactly on the $x$ axis.

Upon increasing $J$ (not shown), care is advised due to the possibility of indirect gap closing. Such gap closings do not alter the topological indices of (complete) individual bands, given that no level crossings are involved. Thus, it must be checked explicitly that the lowest bulk band does not pass through zero energy.

We note that the parameter space is high-dimensional and the diagrams shown here display only a small part of the entire space. However, we have shown that a sizable topological region of signature $(1|0,0,-1)$ exists. Furthermore, the conditions formulated for the continuum model are helpful in navigating to the desired phase in this high-dimensional space, although many features of the tight-binding model are not captured in the continuum limit (which would be recovered for $n\rightarrow\infty$).

\subsection{Majorana bound states at disclinations}\label{Sec:Disclinations}
\begin{figure*}[t]
\includegraphics[width=\textwidth]{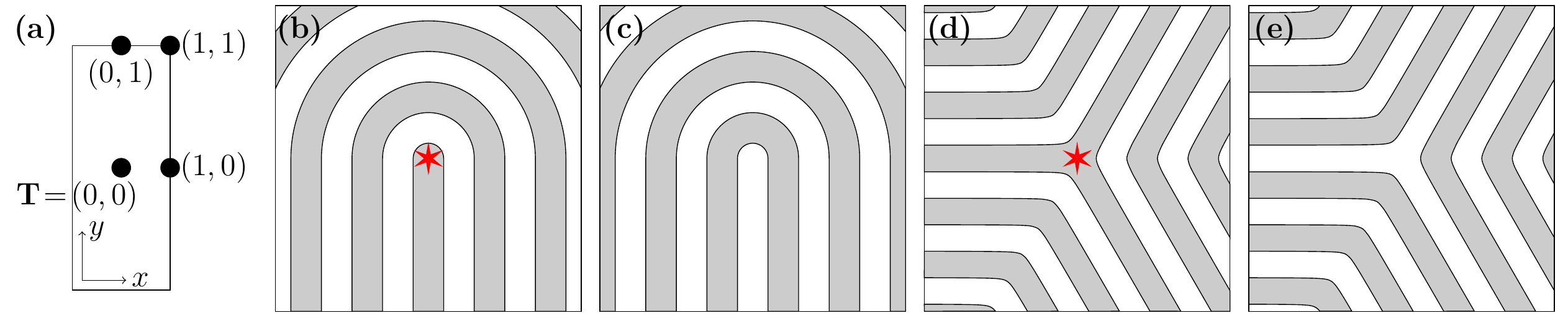}
\caption{\label{Fig:DisclinationTypes}(a) The four possible rotation centers in the real-space unit cell with their respective translation vector $\mathbf{T}\in\mathbb{Z}_2\times\mathbb{Z}_2$ for a disclination in a $C_2$ system. [(b)--(e)] The four types of disclinations in the helimagnet-superconductor hybrid: (b) $\Omega=+\pi$, $T_y=1$, $\Theta=1$; (c) $\Omega=+\pi$, $T_y=0$, $\Theta=0$; (d) $\Omega=-\pi$, $T_y=1$, $\Theta=1$; and (e) $\Omega=-\pi$, $T_y=0$, $\Theta=0$. Shaded regions represent schematically the effective topological wires (see main text); white regions are trivial. In panels (b) and (d), the red star indicates where the disclination Majorana mode is expected --- namely at a wire end or trijunction, respectively. Recall that, in our nomenclature, the $y$ direction is always perpendicular to the stripes.}
\end{figure*}
It is known that symmetry-protected zero-energy states appear not only at boundaries of a topological phase  but also at defects within the bulk \cite{CTS16}. The type of defects that are capable of hosting such bound states depends on the specifications of the topological phase. The fundamental defects of $C_n$ systems are disclinations. They are generally characterized by their Frank angle $\Omega$, describing the lattice rotation, and a translation vector $\mathbf{T}$. The latter essentially specifies the position of the disclination's rotation center in the unit cell. In a $C_2$ symmetric system, $\Omega=\pm\pi$ and there are four inequivalent possibilities to place the rotation center in the unit cell: at the center, a corner, or the center of the edge in either $x$ or $y$ direction; cf. Fig.~\ref{Fig:DisclinationTypes}(a). In total, this yields eight distinct types of disclinations.

According to the theory of topological crystalline superconductors, the Majorana parity at a disclination is given by \cite{TeH13, GTH13, BTH14}
\begin{equation}\label{Eq:MBSCriterion}
\Theta = \inbra{\mathbf{T}\cdot\mathbf{G}_\nu + \frac{\Omega}{2\pi}\inpar{\chern+[X]+[Y]+[M]}} \mod 2,
\end{equation}
where $\mathbf{G}_\nu$ is the vector of \emph{weak} topological invariants. These are related to the crystalline indices through
\begin{equation}
\mathbf{G}_\nu = \begin{pmatrix}[X]+[M] \mod 2\\ [Y]+[M] \mod 2\end{pmatrix}.
\end{equation}%
In all phases that we have found in the previous subsection, $[X]=[M]$. Then Eq.~\eqref{Eq:MBSCriterion} simplifies to
\begin{equation}\label{Eq:MBSCriterionSimple}
\Theta = \inbra{T_y([Y]+[M]) + \frac{\chern+[Y]}{2} + [M]} \mod 2.
\end{equation}
It is reassuring that $T_x$ drops out. We are interested in phenomena that depend only on the magnetic texture and neither on details of the atomic lattice nor on exact commensurability (which in the present model is built in). Indeed, now the Majorana parity is insensitive to a shift of the rotation center in the $x$ direction. Consequently, the number of relevant disclination types is reduced to four. The sign of $\Omega$ is also irrelevant for the presence of MBS but corresponds to different magnetic patterns. We will therefore keep this distinction.

For the $(1|0,0,-1)$ phase, the four disclination types including their expected MBS are shown in Fig.~\ref{Fig:DisclinationTypes}. When the disclination has its center at parallel magnetization and external field ($T_y=1$), we expect a localized MBS. In the opposite case ($T_y=0$), no symmetry-protected state appears.

In the $(1|0,0,-1)$ phase the existence of MBS at disclinations can be understood heuristically in a continuum fashion if one thinks of stripes where $m_z$ and $B$ have the same sign as effective Majorana wires, and stripes where the sign is opposite as trivial regions, similar to Fig.~\ref{Fig:TopView}. In the topological stripes, $\magn$ and $B$ add up and thereby allow for topological band inversion, whereas they annihilate (partially) in the trivial regions. In this wire picture of the system, disclinations with a center inside the topological region correspond to wire ends ($\Omega=+\pi$) or trijunctions ($\Omega=-\pi$). For topological phases with the opposite Majorana parity at disclinations according to Eq.~\eqref{Eq:MBSCriterionSimple}, a possible explanation is that the trivial regions of the continuum model (cf. Fig.~\ref{Fig:TopView}) are so narrow that the hybridization of chiral modes between neighboring topological stripes is stronger than the hybridization of opposite edge modes of the same stripe.

We end this section with two remarks. First, there is an ambiguity in defining the unit cell in real space. Changing this convention would impact both $\mathbf{T}$ and the crystalline indices, but the Majorana parity remains unaffected. Second, it is possible to attach extra flux quanta to the disclinations \cite{BTH14}. This can be implemented by $\hat{r}\rightarrow (-1)^q\hat{r}$ for $q$ flux quanta and thereby changes the sign of $[X]$, $[M]$, and $[Y]$ for odd $q$. Then the Majorana parity at disclinations flips whenever $\chern$ is odd.

\section{Exact diagonalization results}\label{Sec:ExDiag}
\begin{figure*}
\includegraphics[width=\textwidth]{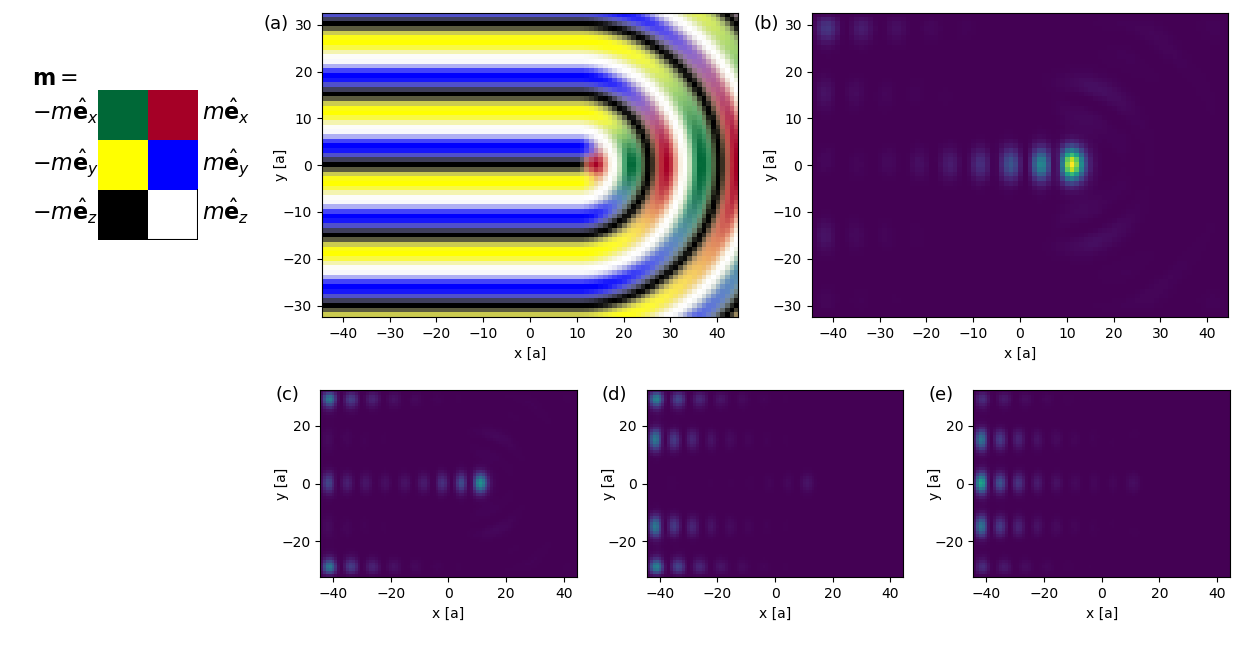}
\caption{\label{Fig:MBS-Stripeend}Exact diagonalization of a disclination with $\Omega=+\pi$ and $\mathbf{T}=(0,1)$ on a grid of size $N_x=89$, $N_y=65$. (a) Orientation of $\magn(\pos)$, displayed using the color scale to the left, (b) MBS localized at the disclination center, [(c)--(e)] the three eigenstates with lowest energies which can be related to Majorana modes in the system, including at the edge. The state shown in panel (b) was obtained from a linear combination of panel (c) and its particle-hole symmetric partner state. The parameters in this calculation were $\Delta=0.25\,t$, $\mu=0.04\,t$, $J=0.22\,t$, $\frac12 B =-0.17\,t$, and $\lambda=0.085\,t$ and the magnetic period length corresponds to $n=15$. The texture is cycloidal.}
\end{figure*}
\begin{figure*}
\includegraphics[width=\textwidth]{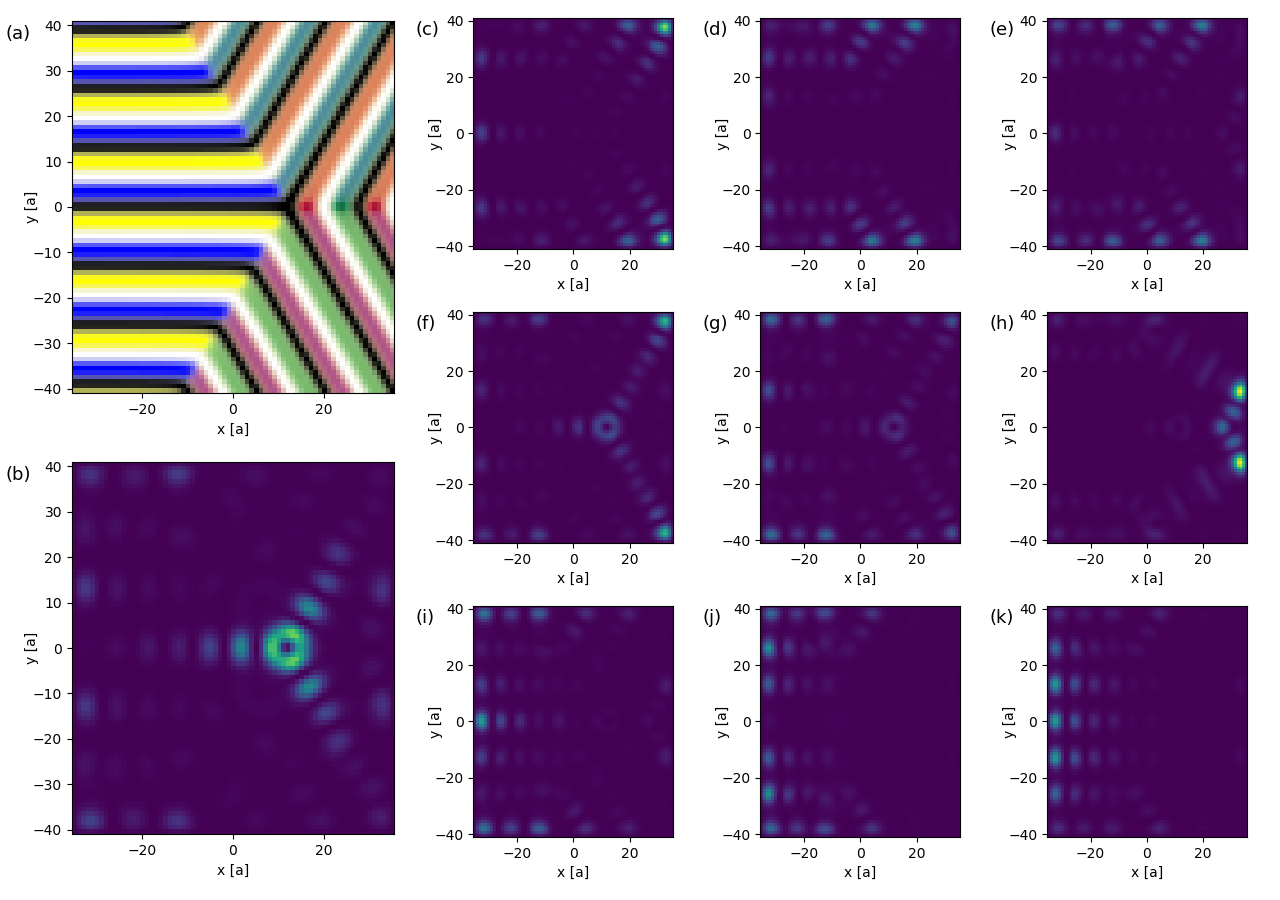}
\caption{\label{Fig:MBS-Trijunction}Exact diagonalization of a disclination with $\Omega=-\pi$ and $\mathbf{T}=(0,1)$ on a grid of size $N_x=71$, $N_y=82$. (a) Orientation of $\magn(\pos)$ with the same color code as in Fig.~\ref{Fig:MBS-Stripeend}, (b) MBS localized at the disclination center, and [(c)--(k)] the nine lowest-energy eigenstates which can be related to Majorana modes in the system, including at the edge. The state shown in panel (b) is a linear combination of panels (f) and (g) and their particle-hole symmetric partner states. The parameters are the same as in Fig.~\ref{Fig:MBS-Stripeend}.}
\end{figure*}
In this section, we back up the findings of the previous section with numerical evidence from exact diagonalization of a tight-binding Hamiltonian on finite two-dimensional systems. Our main finding is the existence of MBS at certain types of disclinations. We verify numerically the existence of such disclination modes for a $+\pi$ disclination with $\mathbf{T}=(0,1)$ in Fig.~\ref{Fig:MBS-Stripeend} and a $-\pi$ disclination with $\mathbf{T}=(0,1)$ in Fig.~\ref{Fig:MBS-Trijunction}. The parameters are chosen such that the system is in the $(1|0,0,-1)$ phase.

For the $+\pi$ disclination, we expect the implemented sample [cf. Fig~\ref{Fig:MBS-Stripeend}(a)] to host three effective Majorana wires. Indeed, the three eigenstates with lowest energies, together with their particle-hole-related partner states (thus corresponding to three true fermionic states) can be related to Majorana modes at the wire ends. These states have most weight at the wire ends and decay along the wires. Taking a linear combination of the lowest-energy state and its particle-hole related partner state, we can isolate a state [Fig~\ref{Fig:MBS-Stripeend}(b)] which is localized at the center of the disclination. The other states lie at the boundary of the system. Given that $\chern=1$, such modes can be expected to form a chiral Majorana band in a sufficiently large system.

For the $-\pi$ disclination, the implemented texture corresponds to nine effective Majorana wires; cf. Fig~\ref{Fig:MBS-Trijunction}(a). Again, we find the correct amount of lowest-energy eigenstates which can consistently be related to Majorana modes. Taking linear combinations of two states and their particle-hole related partners, the disclination MBS can be isolated, cf. Fig~\ref{Fig:MBS-Trijunction}(b). All other states reside at the edge and would again form a chiral band for increasing sample size.

From the results presented in Figs.~\ref{Fig:MBS-Stripeend} and \ref{Fig:MBS-Trijunction}, one can also draw conclusions about the role of the atomic lattice. In the previous section, we have already discussed that the $T_x$ component of the translation vector is irrelevant for the Majorana parity at a disclination. However, perfect alignment of the lattices was always present by design of the Hamiltonian. Here, we emphasize that in our implementation the disclinations involve only the magnetic texture, whereas the atomic square lattice is not rotated at the disclinations. Thus, the atomic lattice is misaligned with the magnetic texture in significant fractions of the area of the simulated samples. The presence of the expected bound states is therefore a clear indication that, in fact, MBS are exclusively determined by the magnetic texture.

In both simulations, one can clearly see that the low-energy wavefunctions spread mainly along the effective wires. This is consistent with the emergent confinement potential, which we have derived within the continuum model.

We note that two-dimensional exact diagonalization is only possible for very limited system sizes. Therefore, the finite-size hybridization of the disclination MBS with the edge modes is comparably strong. Consequently, the energies of our presumed Majorana modes are not zero. The energy of the lowest-lying state is only one order of magnitude below the lowest bulk state. It is reasonable to expect the wavefunction overlap to vanish in a large system, such that the energy of an MBS at a single disclination would converge to zero. Furthermore, it is possible to get a true zero-energy solution by fine-tuning of parameters in such a way that the oscillating tails of the disclination mode and the edge mode interfere destructively. Nevertheless, we decided to present the generic situation without fine-tuning.

\begin{figure*}
\includegraphics[width=\textwidth]{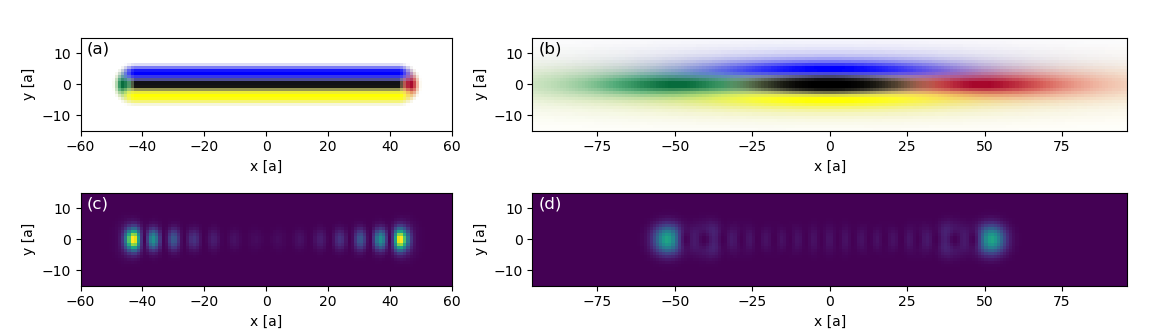}
\caption{\label{Fig:MBS-Stripe-Skyrmion} Magnetic texture of (a) a cycloidal stripe and (b) in an elongated skyrmion of N\'eel type, and the wavefunction weight of the MBS in (c) the stripe and (d) the skyrmion. The parameters in panels (a) and (c) are $\mu=0.04\,t$, $\lambda=0.085\,t$, $J=0.22\,t$, $B=-0.17\,t$, and $\Delta=0.25\,t$ on a $120\times30$ grid. In panels (b) and (d), $\mu=0.05\,t$, $\lambda=0$, $J=0.25\,t$, $B=-0.2175\,t$, and $\Delta=0.25\,t$ on a $192\times30$ grid, where the half-axes of the elliptical skyrmion are $r_x=49$ and $r_y=5$ lattice constants. In particular, note that MBS emerge without Rashba SOC in the skyrmion, which would not be possible in the stripe.}
\end{figure*}

Finally, we discuss the relation of magnetic stripes to elongated skyrmions, which were proposed earlier as effective Majorana wires \cite{GSK18}. A single stripe of the cycloidal phase (i.e., a $2\pi$ N\'eel domain wall with end points) can be interpreted as a limiting case of the elongated skyrmion. In Fig.~\ref{Fig:MBS-Stripe-Skyrmion}, we show both a stripe and a skyrmion with MBS for comparison. There is a qualitative difference, though: In the skyrmion, there is a continuous winding of the spin as one follows the $x$ direction. In the stripe, this winding is zero except at the end points. Therefore, MBS can only be obtained if the Rashba constant $\lambda\neq0$ in the latter case, whereas they can be stabilized by the bare synthetic SOC in the elongated skyrmion. However, as the skyrmion shape gets more eccentric, the $k_x$-related contribution of the SOC would tend to zero. In practice, this limits the length of a skyrmion-based quasi-wire without a Rashba term.

\section{Chiral modes at a domain wall}\label{Sec:DomainWall}
In this section, we consider the hybridization of MBS to chiral modes in the topological emergent-wire phase. It is clear that the Chern number dictates the chirality and amount of protected edge modes \cite{CTS16}. What is not captured by the Chern number itself, though, is the fact that the group velocity of the chiral modes in the $C_2$ system can differ significantly on edges of different orientation with respect to the magnetic texture. This anisotropy of edge modes is apparent within the picture of emergent Majorana wires from the difference in the MBS density on edges in $x$ or $y$ direction. Namely, the density of such MBS will be large on edges that are approximately perpendicular to the effective wires (edges in $y$ direction), corresponding to a small group velocity of a chiral mode. On the other hand, edges approximately oriented along the parallel direction would exhibit a tiny MBS density and thereby a large group velocity. Below, we derive fast and slow modes for the case of a certain domain boundary inside the MSH.

\begin{figure}[t]
\includegraphics[width=\columnwidth]{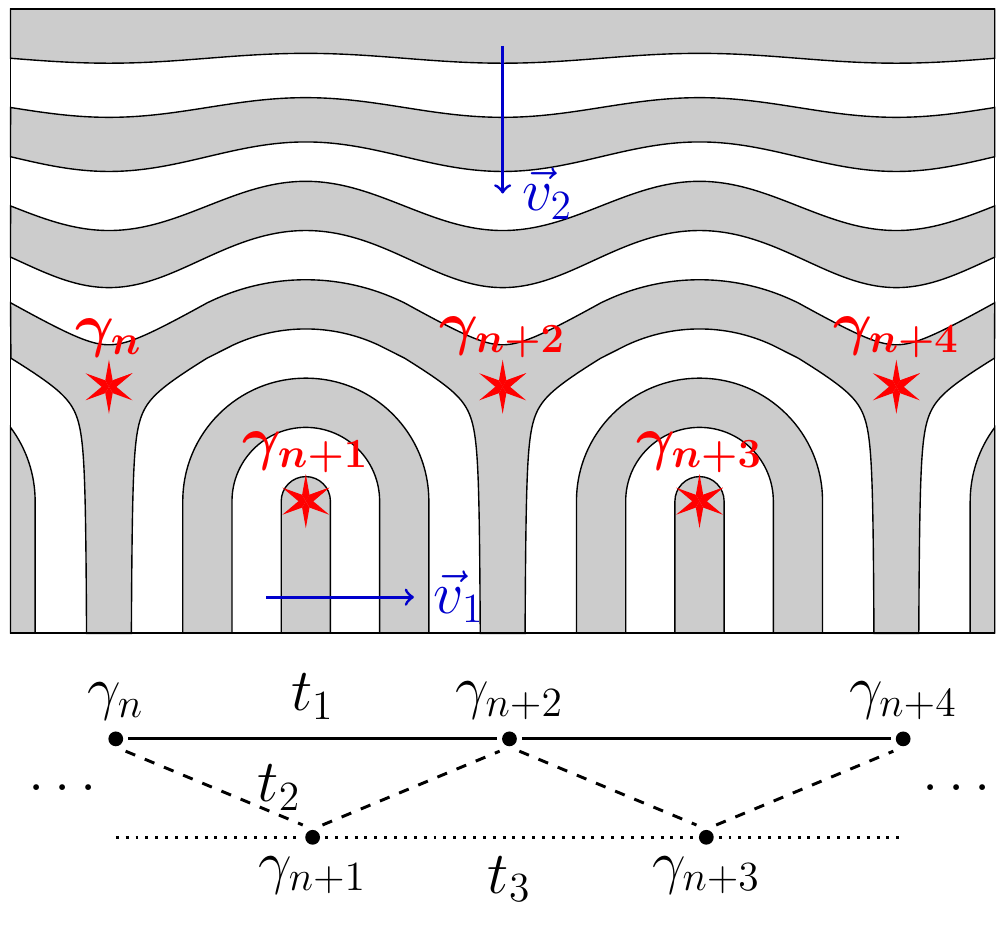}
\caption{\label{Fig:DomainWall}Schematic depiction of a 90-degree domain wall in a helimagnet, where $\vec{v}_1$ and $\vec{v}_2$ denote the orthogonal helix vectors in the two adjacent domains. Here, we assume that the domain wall comprises only topological disclinations (cf. Fig.~\ref{Fig:DisclinationTypes}) and indicate the location of MBS by red stars. Below, the chain of Majorana states with couplings $t_1,t_2,t_3$ is shown.}
\end{figure}

Recently \cite{SMK18}, it was observed experimentally and explained theoretically that domain walls in helimagnets between domains of different helix vector orientations consist of a string of disclination points if the adjacent helix vectors span an angle sufficiently close to $90$ degrees. Our results from the previous sections imply that such a domain wall on a superconductor would bind a MBS at each disclination which is of a proper type. Consequently, a chain of MBS can be expected, such that the system resembles a version of the Kitaev chain. We illustrate this situation for a $90$-degree domain wall with an all-topological string of disclinations in Fig.~\ref{Fig:DomainWall}.

Given that the domain-wall MBS are relatively close to each other (a few magnetic period lengths apart), their overlap would lead to non-negligible coupling terms. As the wavefunctions spread mainly along the effective wires, similar to, e.g., Fig.~\ref{Fig:MBS-Stripe-Skyrmion}(c), one can expect that the coupling will be strongest between Majorana modes at $-\pi$ disclinations (even sites in Fig.~\ref{Fig:DomainWall}). To couple the other modes, the wavefunctions have to span across multiple magnetic stripes, thus penetrating several barriers of the emergent confinement. These couplings will likely be significantly smaller. With the coupling constants $t_1$ between $-\pi$-disclination MBS, $t_2$ between MBS at neighboring $+\pi$ and $-\pi$ disclinations, and $t_3$ between $+\pi$-disclination MBS (cf. the lower part of Fig.~\ref{Fig:DomainWall}) we expect that $|t_1|>|t_2|>|t_3|$.

\begin{figure}[t]
\includegraphics[width=\columnwidth]{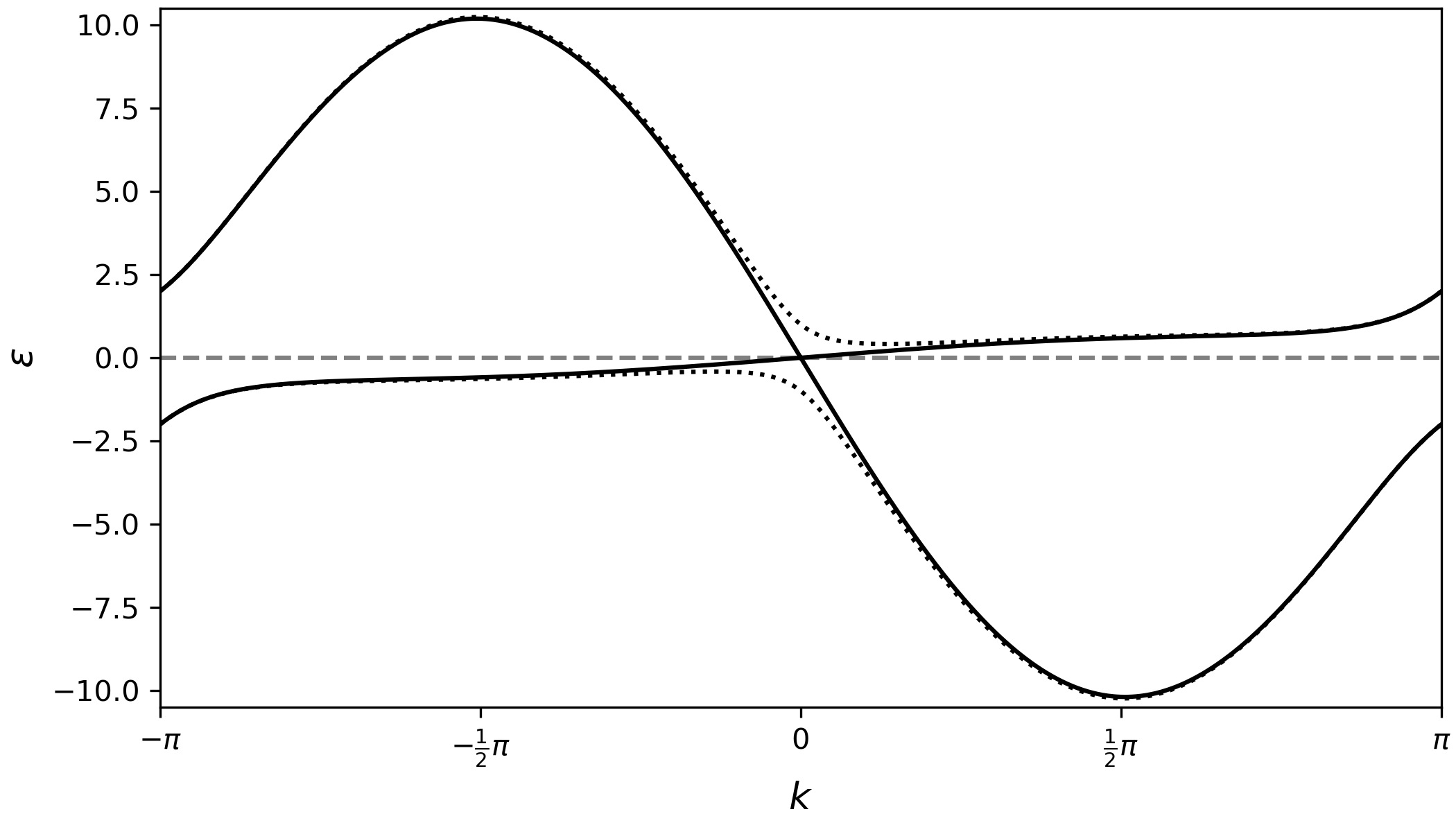}
\caption{\label{Fig:DW-dispersion}The dispersion of the modes at a 90-degree domain wall. They result from the hybridization of disclination MBS, forming a fast mode and a slow mode, which can be related to the edge modes of the adjacent domains. Here, the coupling constants are $t_1=5$, $t_2=1$, and $t_3=-0.2$. The dotted lines indicate the gap opening at $k=0$ if staggering of size $(\Delta t_2)=1$ is introduced in $t_2$.}
\end{figure}

Using the index $j$ to count the disclinations, the Hamiltonian of this chain is
\begin{eqnarray}
H &=& - it_1\sum_j \gamma_{2j}\gamma_{2(j+1)} -it_2\sum_{j}\gamma_j\gamma_{j+1} \notag\\ &&
      - it_3\sum_j \gamma_{2j+1}\gamma_{2(j+1)+1}\,,
\end{eqnarray}
with Majorana operators $\gamma_j$. Pairs of MBS at neighboring $+\pi$ and $-\pi$ disclinations can be combined into fermion modes, $c_{2j}=\frac12\inpar{\gamma_{2j}-i\gamma_{2j+1}}$. Furthermore, we take a Fourier transform, $c_{2j}=\frac{1}{\sqrt{N}}\sum_k \exp\inpar{-i[2j]k}c_k$, and write the reciprocal-space Hamiltonian in BdG form as $H=\frac{1}{N}\sum_k\Psi_k^\dagger \mathcal{H}(k)\Psi_k$ with $\Psi_k=(c_k, c^\dagger_{-k})$. Then,
\begin{eqnarray}\label{Eq:BdG-DW}
\mathcal{H}(k) &=& t_2\inpar{1-\cos k}\tau_z - \inpar{t_1+t_3}\sin k\notag\\
&&{} + t_2\sin k \,\tau_y - \inpar{t_1-t_3}\sin k \,\tau_x\,,
\end{eqnarray}
and the eigenenergies of the system are
\begin{eqnarray}
\varepsilon_{1,2} &=& -(t_1+t_3)\sin k \notag\\&&
\pm \sqrt{2t_2^2\inpar{1-\cos k} + \inpar{t_1-t_3}^2\sin^2k}\,.\label{Eq:DWdispersion}
\end{eqnarray}
The dispersion of the modes is displayed in Fig.~\ref{Fig:DW-dispersion}.
We observe that the pronounced asymmetry of $t_1$ and $t_3$ causes the formation of two modes with different velocities: a fast mode and a (nearly flat) slow mode. As expected, we can link these modes to the edge modes of the adjacent domains, where the domain wall is parallel with regard to one domain and perpendicular with regard to the other domain. Given that the Chern number, i.e., the edge mode chirality, is the same on both sides of the domain wall, we infer that the fast and the slow modes must run in opposite directions. Hence, $t_1$ and $t_3$ will have opposite signs.

It is clear from Eq.~\eqref{Eq:DWdispersion} that the domain-wall Hamiltonian is generically gapless at $k=0$. In that sense, it corresponds to a Kitaev chain exactly at the transition point. In order to open a gap around zero energy, one would have to introduce additional staggering in the nearest-neighbor coupling, $t_2\rightarrow t_2\pm\frac12(\Delta t_2)$. The modified dispersion relation would then become
\begin{widetext}
\begin{equation}
\varepsilon_{1,2} = -(t_1+t_3)\sin k \pm 
\sqrt{2t_2^2\inpar{1-\cos k} + \frac{(\Delta t_2)^2}{2}\inpar{1+\cos k} + \inpar{t_1-t_3}^2\!\sin^2k},
\end{equation}
\end{widetext}
such that $\varepsilon_{1,2}(k=0)=\pm|\Delta t_2|$. This case is also indicated in Fig.~\ref{Fig:DW-dispersion}. Only then could the domain wall itself turn topological, as in Kitaev's model. Staggering would probably not appear spontaneously, in contrast to the domain wall itself, because the potential energetic benefit of the nearly zero-energy states is negligible compared to the energy scale of the magnetic interaction terms. As an example, in the Co/Ru(0001) candidate system \cite{HDL18}, the direct exchange energy is $13\,\text{meV}$, while the zero-temperature $s$-wave gap of the substrate is only $0.07\,\text{meV}$ \cite{HuG57} --- which is still at least one order of magnitude larger than the energy of bound states inside the effective topological gap (cf. Figs.~\ref{Fig:Phases-B-mu} and \ref{Fig:Phases-lambda-n}).

A real domain wall will exhibit a significant amount of disorder \cite{SMK18}. For instance, the disclinations may not be strictly equidistant. The gapless Kitaev chain beyond nearest-neighbor coupling in the presence of both disorder and interactions represents an interesting theoretical model with rich behavior, as demonstrated in a recent paper \cite{KSM19}. We suggest that 90-degree helimagnet domain walls (and likely also cycloidal domain walls) on a superconducting substrate may open a possibility to make such models also accessible in experiments.

As a possible limitation of domain-wall Majorana chains, only some of the involved disclinations may have odd Majorana parity (cf. Fig~\ref{Fig:DisclinationTypes}) in reality. Thus, a domain wall in the magnetic texture will not always induce separate domains of the effective topological superconductor with boundary modes in between. If MBS appear sparsely along the domain wall, one may rather view them as a set of statistically placed isolated modes instead of a chiral chain.

\section{Conclusion}\label{Sec:Conclusion}
We employed three different methods to investigate topological phases in a superconducting film coupled to a chiral magnet with helical or cycloidal order, namely a continuum model approach, tight-binding calculations based on the magnetic unit cell, and exact diagonalization of finite systems. Each of these calculations independently suggests that there is a phase in which parallel effective Majorana wires emerge in consequence of the magnetic texture.

The continuum approach in Sec.~\ref{Sec:Continuum} allowed us to derive conditions for this topological phase, taking into account corrections to the chemical potential and the SOC resulting from a local spin-space rotation according to the orientation of the net exchange field. In particular, the effective chemical potential evokes a spatial confinement and thereby the formation of wires. The total SOC consisting of both Rashba and synthetic contributions is, in general, anisotropic in momentum space. Our analysis of gap closings showed that a full bulk gap demands nonzero Rashba SOC and that the gap will generically exhibit weak spots in momentum space. Furthermore, we recovered the well-known band-inversion condition at $k=0$. Notably, though, exchange fields that are too large can cause an indirect closure of the gap.

Based on the tight-binding model of Sec.~\ref{Sec:C2TB}, we have established the superconducting hybrid system with spiral magnetic order as an example of a $C_2$-symmetric topological crystalline superconductor in two dimensions. We found that a cycloidal magnetic texture allows for a larger topological gap than a helical texture, in agreement with the continuum model. The hallmark of the topological phase is the existence of localized MBS at $\pm\pi$ disclinations, if their rotation center is placed suitably with respect to the magnetic unit cell. Such disclinations correspond to ends or trijunctions of Majorana wires within the continuum picture, where one would also expect MBS. We have confirmed this key result numerically in Sec.~\ref{Sec:ExDiag} by exact diagonalization of systems with disclinations on finite two-dimensional lattices, cf. Figs.~\ref{Fig:MBS-Stripeend} and \ref{Fig:MBS-Trijunction}. We have also verified the emergent-wire interpretation for a single 360-degree magnetic domain wall and compared it to an elongated skyrmion.

Finally, we have discussed a 90-degree domain wall between domains of differently oriented magnetic spirals (cf. Sec.~\ref{Sec:DomainWall}), which is formed by a string of disclinations. The case where all disclinations have odd Majorana parity turned out to be an interesting example for the hybridization of the disclination MBS into chiral modes, where we have found two counterpropagating modes of different velocity along the domain wall. We discussed the relation of these modes to the chiral boundary modes of the topological phase with Chern number 1.

All magnetic structures that we studied in this paper can appear spontaneously in chiral magnets without a need for further nano-engineering beyond the creation of the magnet-superconductor interface. In addition, helical and cycloidal textures, including disclination defects, may be found more frequently than the previously discussed skyrmions \cite{YSK16,GSK18,RGM19,GMS19}. The impact of induced superconductivity on the properties of chiral magnets is largely unexplored experimentally, though.

We hope that our work will prepare the ground for further theoretical and experimental progress on Majorana physics. Future work may address MBS at defects in other periodic magnetic textures, e.g., the skyrmion lattice \cite{MKS15,MBG20}. Furthermore, the helical or cycloidal MSH can be inhomogeneous in many ways apart from the cases studied in this work. Randomly positioned disclinations or dislocations (i.e., disclination pairs) can appear and could potentially lead to more complex effective Majorana wire networks. Thus, our proposal may open a path to study such networks in experiments without a need to fabricate actual nanowires. Close to the transition to skyrmion phases, mixtures of skyrmions and magnetic stripes of different orientations and lengths may form. Thus, two-dimensional systems with numerous MBS configurations can be envisioned, including mixtures of disclination MBS and skyrmion-induced MBS \cite{YSK16, GSK18, RGM19, GMS19}. In such settings, dynamical properties of the MSH subject to charge or spin currents as means to control the MBS remain to be investigated. For instance, helimagnets can be manipulated by means of a weak electric current \cite{MYK20}.

\begin{acknowledgments}
We thank Markus Garst and Wulf Wulfhekel for inspiring discussions on this project. The work was supported by the Deutsche Forschungsgemeinschaft via the Grants No. MI 658/12-1 (joint DFG-RFBR project) and No. MI 658/13-1 (joint DFG-RSF project). I.V.G. acknowledges support by the Russian Science Foundation through Grant No. 17-12-01182~c.
\end{acknowledgments}

\appendix

\section{Spin-alignment transformation with a general field}\label{App:SAT}
Here we list the result of the SAT applied to the continuum Hamiltonian for a general field $\Magn(\pos)$ in two dimensions, assuming the field varies sufficiently slowly such that $k_{x,y}$ are locally good quantum numbers. The field orientation is given by $f(\pos)$ and $g(\pos)$ as in the main text, but we suppress the argument $\pos$ below. For the transformation of the kinetic term $h_\text{kin}=\hbar^2 k^2/(2m)$, we find
\begin{equation}
\tilde{h}_\text{kin} = 
h_\text{kin} - \mu_\text{kin} + h_\text{syn}^\text{so}\,,
\end{equation}
including a correction to the chemical potential
\begin{equation}
\mu_\text{kin} = -\frac{\hbar^2}{8m}\inbra{(\nabla g)^2 + 2(\nabla f)^2(1-\cos g)}
\end{equation}
and the synthetic spin-orbit coupling
\begin{equation}\label{Eq:SOsyn}
h_\text{syn}^\text{so} = \frac{\hbar k_y}{2m}
\inbra{(\nabla g)\sigma_a + (\nabla f)\inpar{\sin g\, \sigma_b + [1-\cos g]\sigma_z}},
\end{equation}
where we have used the auxiliary Pauli matrices
\begin{eqnarray}
\sigma_a &=& \sigma_x \sin f - \sigma_y \cos f\,, \\
\sigma_b &=& \sigma_x \cos f + \sigma_y \sin f\,.
\end{eqnarray}

For the Rashba term $h_\text{R}=\hbar\alpha(\sigma_xk_y - \sigma_yk_x)$, the transformation results in
\begin{equation}
\tilde{h}_\text{R} = h_\text{R} + h^\text{so}_\text{twist} - \mu_\text{R},
\end{equation}
with a \quotmarks{twisted} spin-orbit contribution
\begin{equation}
h^\text{so}_\text{twist} = \hbar\alpha\inbra{\sigma_b(\cos g - 1)-\sigma_z\sin g}\inpar{k_y\cos f - k_x\sin f}
\end{equation}
and another correction to the chemical potential,
\begin{eqnarray}
\mu_\text{R} &=&
  -\frac{\hbar\alpha}{2}\inbra{\inpar{\partial_y g}\sin f + \inpar{\partial_x g}\cos f } \notag\\&&
  -\frac{\hbar\alpha}{2}\inbra{\inpar{\partial_y f}\cos f - \inpar{\partial_x f}\sin f}\sin g.
\end{eqnarray}
In total, the effective chemical potential is
\begin{equation}
\mu_\text{eff} = \mu + \mu_\text{kin} + \mu_\text{R}
\end{equation}
and the overall SOC reads
\begin{equation}
h^\text{so}_\text{tot} = h^\text{so}_\text{syn} + h_\text{R} + h^\text{so}_\text{twist}.
\end{equation}
As mentioned in the main text, the singlet pairing term is invariant under the SAT.

On a technical note, transforming the momentum-dependent terms straightforwardly leads, at first, to imaginary magnetic fields appearing along with non-Hermitian (not properly symmetrized) spin-orbit terms. It is only after the commutation of all terms into a symmetric form that the unphysical imaginary fields drop out. This has also been noted earlier \cite{KWF12}.

\section{Terms of the tight-binding Hamiltonian}\label{App:TB}
The tight-binding Hamiltonian based on magnetic unit cells reads, in momentum space,
\begin{equation}
H_\text{t.b.}(\mom) = \sum_{\boldsymbol{\delta}} e^{i\mom\cdot\boldsymbol{\delta}}H_{\boldsymbol{\delta}}\,,
\end{equation}
where $\boldsymbol{\delta}=0$ for terms within one unit cell, whereas $\boldsymbol{\delta}=\pm a\unitvec_x \pm na\unitvec_y$ for the four hopping terms to the neighboring unit cells. We define the following matrices in the $n\times n$-dimensional space of internal sites: $\mathbb{I}_i$ has the element $1$ at the $i$th position on the diagonal as its only non-zero entry, $\mathbb{D}_{\pm}$ denote the upper ($+$) and lower ($-$) secondary diagonals, $\mathbb{U}$ has the only nonzero entry $1$ in the upper right corner, and $\mathbb{L}$ has the only nonzero entry $1$ in the lower left corner. Then the contributions $H_{\boldsymbol{\delta}}$ can be expressed in terms of the atomic $4\times 4$ blocks 
\begin{eqnarray}
H_\text{on-site}^i &=& -\mu\tau_z + \Magn_i\cdot\tilde{\sigmab}\tau_z - \Delta\tau_y\sigma_y\,, \\
H_{\pm x} &=& -t\tau_z \pm \frac{i\lambda}{2}\sigma_y\tau_z\,, \\
H_{\pm y} &=& -t\tau_z \mp \frac{i\lambda}{2}\sigma_x
\end{eqnarray}
as follows:
\begin{equation}
H_0 = \sum_{i=1}^n H_\text{on-site}^{i}\otimes \mathbb{I}_i + H_{-y}\otimes\mathbb{D}_{+} + H_{+y}\otimes\mathbb{D}_{-}\,,
\end{equation}
\begin{equation}
H_{\pm a\unitvec_x} = H_{\pm x} \otimes \unity_{(n\times n)}\,,
\end{equation}
\begin{equation}
H_{  + na\unitvec_y} = H_{+y} \otimes \mathbb{U}\,,
\end{equation}
\begin{equation}
H_{  - na\unitvec_y} = H_{-y} \otimes \mathbb{L}\,.
\end{equation}
The Pauli vector $\tilde{\sigmab}$ is the same as in Sec.~\ref{Sec:SAT}. The total exchange field at each internal site is given by
\begin{eqnarray}
\Magn_i &=&
\inbra{\frac12 g\mu_BB + J|\magn|\cos\inpar{2\pi\frac{i-\frac12}{n}}}\unitvec_z \notag\\
&&{}+ J|\magn|\sin\inpar{2\pi\frac{i-\frac12}{n}}\unitvec_j
\end{eqnarray}
similar to Eq.~\eqref{Eq:MagTexture}, with $i=1\hdots n$ and $j=x,y$ for helical or cycloidal magnets.

\bibliography{MBS-disclination-v5}

\end{document}